\documentclass[12pt]{article}

\usepackage{graphicx}
\usepackage{amsmath,amssymb}
\usepackage{amsthm,amscd,mathrsfs}

\newcommand{\be}{\begin{eqnarray}} 
\def\ne{\nonumber\end{eqnarray}}
\newcommand{\ee}{\end{eqnarray}}
\newcommand{\nn}{\nonumber}
\def\l{\left}
\def\r{\right}
\def\hf{{1\over2}}
\def\d{\partial}
\def\cd{\nabla}
\def\Det#1#2{{\rm Det}_{(#1)}\left[{#2}\right]}

\newcommand{\gsim}{\mathrel{\mathop{\kern-0pt >}\limits_{\sim}}}
\newcommand{\lsim}{\mathrel{\mathop{\kern-3pt <}\limits_{\sim}}}
\def\sitarel#1#2{\mathrel{\mathop{\kern-0pt #1}\limits_{#2}}}
\def\inbar{\,\vrule height1.5ex width.4pt depth0pt}
\def\IC{\relax\hbox{$\inbar\kern-.3em{\rm C}$}}
\def\IH{\relax{\rm I\kern-.18em H}}
\def\IP{\relax{\rm I\kern-.18em P}}
\def\IQ{\relax\hbox{$\inbar\kern-.3em{\rm Q}$}}
\def\IR{\relax{\rm I\kern-.18em R}}
\font\cmss=cmss10 \font\cmsss=cmss10 at 7pt
\def\IZ{\relax\ifmmode\mathchoice
{\hbox{\cmss Z\kern-.4em Z}}{\hbox{\cmss Z\kern-.4em Z}}
{\lower.9pt\hbox{\cmsss Z\kern-.4em Z}}
{\lower1.2pt\hbox{\cmsss Z\kern-.4em Z}}\else{\cmss Z\kern-.4em Z}\fi}
\def\1{{\bf 1}}

\def\G{\Gamma}
\def\g{\gamma}
\def\s{\sigma}
\def\dl{\delta}
\def\La{\Lambda}
\def\la{\lambda}
\def\e{\epsilon}
\def\ve{\varepsilon}
\def\vp{\varphi}
\def\a{\alpha}
\def\b{\beta}
\def\da{\dot{\alpha}}
\def\db{\dot{\beta}}
\def\om{\omega}
\def\N{{\cal N}}
\def\L{{\cal L}}
\def\bz{{\bar{z}}}
\def\bzz{{\bar{z}z}}
\def\tr#1{{\rm tr}{\left[{#1}\right]}}
\def\t#1{\tilde{#1}}
\def\as{{\left(\alpha\cdot\sigma\right)}}

\def\mt{\tilde{m}}
\def\lmm{{l,m,\tilde{m}}}
\def\t{\tilde}
\def\sumlmm{\sum_{l=0}^{\infty}\sum_{m=-l}^{l}\sum_{\tilde{m}=-l}^{l}}

\begin{document}

\vspace*{-1.2cm}
\hfill{UT-12-33}\\

\vskip 1cm
\centerline{\Large\bf 
5D SYM and 2D $q$-Deformed YM
}

\vskip .5in

\centerline{\sc Yasutaka Fukuda, Teruhiko Kawano, ~and~ Nariaki Matsumiya}

\vskip .1in
\centerline{\it Department of Physics, University of Tokyo, 
Hongo, Tokyo 113-0033, Japan}

\vskip .5in
{\small 
We study the AGT-like conjectured relation 
of a four-dimensional gauge theory on $S^3\times{S}^1$ 
to two-dimensional $q$-deformed Yang-Mills theory on a Riemann surface $\Sigma$ 
by using a five-dimensional $\N=2$ supersymmetric Yang-Mills theory 
on ${S}^3\times\Sigma$, following the conjectured relation 
of the six-dimensional ${\cal N}=(2,0)$ theory on $S^1$ to 
the five-dimensional Yang-Mills theory. 
Our results are in perfect agreement with both of the conjectures. 
}

\vskip .3in
\setcounter{footnote}{0}
\section{Introduction}

The authors of the paper \cite{Rastelli} have discussed 
that the partition function (the superconformal index) 
of a four-dimensional ${\cal N}=2$ gauge theory on $S^3\times{S}^1$ 
is given by two-dimensional $q$-deformed Yang-Mills theory 
on a Riemann surface $\Sigma$ in the zero area limit. 
This is analogous to the Alday-Gaiotto-Tachikawa relation \cite{AGT} 
of a four-dimensional ${\cal N}=2$ gauge theory on $S^4$ and 
two-dimensional Liouville theory on a Riemann surface $\Sigma$. 
Following the prescription \cite{Gaiotto} of Gaiotto, the four-dimensional 
gauge theories are specified by the corresponding Riemann surface $\Sigma$, 
and it is widely thought that the putative 
six-dimensional ${\cal N}=(2,0)$ theories 
on $S^3\times{S}^1\times\Sigma$ and $S^4\times\Sigma$ 
underlie behind the former relation and the latter one, respectively.  

However, the six-dimensional ${\cal N}=(2,0)$ theory has not yet  
been formulated as a full-fledged theory. 
Therefore, one cannot use the ${\cal N}=(2,0)$ theory 
to check the relations directly. But, it has been argued in 
\cite{Douglas, Lambert} that a five-dimensional ${\cal N}=2$ Yang-Mills theory 
itself yields the full ${\cal N}=(2,0)$ theory on $S^1$. 
If this is the case, we can study the proposal of \cite{Rastelli} 
more directly by using the five-dimensional ${\cal N}=2$ Yang-Mills theory 
on $S^3\times\Sigma$. In this paper, we will carry out the localization 
procedure in the five-dimensional theory to seek the relation of it with 
the two-dimensional $q$-deformed Yang-Mills theory on the Riemann surface. 

In the previous paper \cite{flatvector}, two of us have already studied 
the partition function of 
the five-dimensional ${\cal N}=1$ Yang-Mills theory on $S^3\times\IR^2$ 
by using localization\footnote{See 
\cite{KZ,HosomichiTerashima,KK,KKL,Terashima,KZ2,KMNZ,JP,Imamura,KL} 
for recent related works on five-dimensional supersymmetric gauge theories.}, 
and found that it yields two-dimensional bosonic 
Yang-Mills theory on $\IR^2$. On the flat $\IR^2$, it was not possible to 
distinguish between the ordinary Yang-Mills theory and the $q$-deformed one.  

In this paper, therefore our previous results on the $S^3\times\IR^2$ 
in \cite{flatvector} 
will be extended to the $\N=1$ theory on  $S^3\times\Sigma$ with 
$\Sigma$ a closed Riemann surface. It will be further extended 
to the $\N=2$ theory by introducing a hypermultiplet into the $\N=1$ theory. 
We will compute the partition function 
in the resulting five-dimensional $\N=2$ theory 
by carrying out the localization method, and will find that 
it is identical to the partition function of 
the two-dimensional $q$-deformed Yang-Mills theory. 
We will see that the parameter $q$ derived from the five-dimensional theory 
is given in terms of the gauge coupling constant and the radius of the round 
$S^3$, and it is in perfect agreement with the prediction from the conjecture 
of \cite{Douglas,Lambert} on the six-dimensional $\N=(2,0)$ theory. 

In the next section, we will give a very brief review on the 
quantization of the two-dimensional $q$-deformed Yang-Mills theory 
on a closed Riemann surface $\Sigma$, and a 
five-dimensional supersymmetric Yang-Mills theory on a flat Euclidean space 
$\IR^5$ will be given in section \ref{flatcase}. 

In order to put the five-dimensional theory on $S^3\times\Sigma$, 
we will explain the Killing spinors on the round $S^3$ 
and the ``partial twisting'' \cite{Bershadsky,Wallcrossing} 
on the surface $\Sigma$ 
to define a supersymmetry transformation on the $S^3\times\Sigma$ 
in section \ref{SUSYParameter}, and 
give the supersymmetry transformations, their algebra, and the Lagrangian 
on the $S^3\times\Sigma$ in section \ref{SUSYalg}. 

In section \ref{Localization}, we will carry out the localization 
procedure to compute the partition function of the five-dimensional theory. 
It will turns out that the result of summing up the one-loop determinants 
yields the partition function of the two-dimensional 
$q$-deformed Yang-Mills theory, 
but not of the two-dimensional ordinary Yang-Mills theory.  

In section \ref{Discussions}, we will discuss that 
the parameter $q$ calculated in the five-dimensional theory 
is in perfect agreement with the prediction of the conjecture 
\cite{Douglas,Lambert} for the six-dimensional $\N=(2,0)$ theory.

In appendix \ref{notations}, our notations on the gamma matrices will 
be explained. Since we will need the Clebsch-Gordan coefficients in the 
one-loop calculation in section \ref{Localization}, 
the Clebsch-Gordan coefficients necessary for the calculations are 
listed in appendix \ref{ClebschGordan}. 

\section{Two-Dimensional $q$-Deformed Yang-Mills Theory on $\Sigma$}
\label{2DYM}

We begin with a very brief review on two-dimensional Yang-Mills theory 
on a closed Riemann surface $\Sigma$. For more details, see \cite{Lecture,qYM}.
The two-dimensional Yang-Mills theory with the gauge group\footnote{
We assume that the gauge group $G$ is simple.} $G$ 
on the Riemann surface $\Sigma$ is described by gauge fields $v_z$, 
$v_\bz$ with the Lagrangian\footnote{
The Lagrangian is the conventional Yang-Mills Lagrangian 
multiplied by a factor two, in our convention.}
\be
\tr{\l(g^{\bzz}v_{\bzz}\r)^2},
\ne
where the field strength $v_{\bzz}$ is defined by 
$\d_\bz{v}_z-\d_z{v}_\bz+ig\l[v_\bz,\,v_z\r]$, and $g^{\bzz}$ is 
the K\"ahler metric on $\Sigma$.

Introducing a scalar field $\phi$ in the adjoint representation of $G$, 
one may rewrite the Lagrangian into
\be
\L_{YM}=\tr{\phi^2-2i\phi\,{g}^{\bzz}v_{\bzz}}.
\ne

Let us consider the quantization of this system by the path integral. 
The first term in the Lagrangian will be treated as a perturbation, 
and the second term is the bosonic part of the Lagrangian of a topological 
field theory. 
We will first perform the path integral over the gauge fields. 
To this end, by using a gauge transformation
\be
\phi\to\phi+ig\l[\om(z,\bz),\phi\r], 
\qquad
v_z\to{v}_z-D_z\om(z,\bz),
\label{2dgaugetrf}
\ne
we impose the gauge fixing condition on the scalar field $\phi$, 
\be
\phi(z,\bz)=\sum_{i=1}^{r}\phi^i\,{}H_i, 
\label{2dgauge}
\ee
where $H_i$ ($i=1,\cdots,r$) are the generators of 
the Cartan subalgebra of $G$ of rank $r$. 
Integrating over the Cartan part of the fluctuations 
\be
\t{v}_z(z,\bz)=\sum_{i=1}^{r}\t{v}^i_z(z,\bz)\,{}H_i,
\ne
of the gauge fields, one obtains the delta functions imposing 
the conditions
\be
\d_z\phi^i=0,
\qquad
\d_\bz\phi^i=0,
\ne
on $\phi$, and they require that $\phi^i$ should be a constant. 
Then, the remaining fluctuations $\tilde{v}_z$, $\tilde{v}_\bz$ 
of the gauge fields, which should be integrated about 
the classical solution $\phi^i={\rm const.}$, are 
\be
\tilde{v}_z=\sum_{\a\in\La}\tilde{v}_z^{\a}(z,\bz)E_{\a},
\ne
where $\La$ is the set of all the roots 
of the Lie algebra of $G$, 
and the root generators $E_{\a}$ satisfy the algebra
\be
\l[H_i,\,E_{\a}\r]=\a_i\,{E}_\a, 
\qquad
\l[E_\a,\,E_{-\a}\r]=\sum_{i=1}^{r}\a_i\,{H}_i\equiv\a\cdot{H},
\ne
with the normalization 
\be
\tr{H^iH^j}=\dl^{i,j},
\qquad
\tr{E_{-\a}E_\a}=1. 
\ne
Therefore, the Lagrangian $\L_{YM}$ gives
\be
\L_{\rm cl}-\sum_{\a\in\La}2g\l(\a\cdot\phi\r)
{g}^{\bzz}\tilde{v}_\bz^{-\a}\tilde{v}_z^{\a},
\ne
with $\l(\a\cdot\phi\r)=\sum_{i=1}^{r}\a^i\phi^i$, 
where $\L_{\rm cl}$ is the classical 
value of the Lagrangian $\L_{YM}$ given by 
\be
\sum_{i=1}^{r}\l[\phi^i\phi^i-2ig^{\bzz}v_{\bzz}^i\,\phi^i\r], 
\ne
with the non-trivial first Chern class 
$\int_{\Sigma}v^i_{z\bz}dz\wedge{d}\bz\not=0$. 

We will add the gauge-fixing term and the ghost term 
\be
\sum_{\a\in\La}\l[b^{-\a}\phi^{\a}-2g\l(\a\cdot\phi\r)\bar{c}^{-\a}c^{\a}\r],
\ne
with the auxiliary fields and the ghosts
\be
b(z,\bz)=\sum_{\a\in\La}b^{\a}(z,\bz)E_{\a},
\qquad
c(z,\bz)=\sum_{\a\in\La}c^{\a}(z,\bz)E_{\a},
\qquad
\bar{c}(z,\bz)=\sum_{\a\in\La}\bar{c}^{\a}(z,\bz)E_{\a},
\ne
to impose the gauge fixing condition (\ref{2dgauge}) correctly. 

The integration over the fields $b^{\a}$ and $\phi^{\a}$ is trivial, 
while the integration over the fluctuation of the gauge fields and 
the pair of ghosts gives 
\be
\prod_{\a\in\La}{\Det{0,0}{2g\l(\a\cdot\phi\r)}\over
{\Det{1,0}{2g\l(\a\cdot\phi\r)}}},
\ne
where $\Det{p,q}{D}$ is the functional determinant of the operator $D$ 
over the space of the $(p,q)$-forms on the Riemann surface $\Sigma$.
As explained in \cite{Lecture}, due to the Hodge decomposition, it yields 
\be
\prod_{\a\in\La}\l[2g\l(\a\cdot\phi\r)\r]^{\chi(\Sigma)/2}
=\prod_{\a\in\La_+}\l[2ig\l(\a\cdot\phi\r)\r]^{\chi(\Sigma)},
\label{Z2DYM}
\ee
up to an overall constant, 
with $\Lambda_+$ the set of the positive roots, 
where $\chi(\Sigma)$ is the Euler character of $\Sigma$.

Let us proceed to the $q$-deformed Yang-Mills theory by making 
the scalar $\phi^i$ periodic. Following \cite{qYM}, 
we will use the method of images to extend (\ref{Z2DYM}) to 
this case. For brevity, we will take the $SU(2)$ gauge group, 
and the measure (\ref{Z2DYM}) is replaced via the method 
of images by 
\be
\prod_{n=-\infty}^{\infty}\l(2\sqrt{2}ig\phi+i{n\over{l}}\r)^{\chi(\Sigma)}
=\l[2i\sin\l({2\sqrt{2}{\pi}gl\phi}\r)\r]^{\chi(\Sigma)},
\label{ZqYM}
\ee
for the periodicity $\phi \to \phi+n/(2\sqrt{2}{g}l)$, 
with the abbreviation $\phi=\phi^1$ here. 
Note that the zeta regularization has been used.

Since the classical Lagrangian $\L_{\rm cl}$ contains 
the term $-4\sqrt{2}\pi{m}\phi/g$ from the non-trivial flux 
\be
{1\over2\pi}\int_\Sigma v_{\bzz}d\bz\wedge{dz}={\sqrt{2}\over{g}}m, 
\qquad \l(m \in \IZ\r) 
\ne
whose normalization will be explained in section \ref{Discussions}, 
summing over all the fluxes $m$ reduces the integration over $\phi$ 
into the summation over $\phi={ing}/(2\sqrt{2})$ with $n\in\IZ$. 
Therefore, at each point $\phi={ing}/(2\sqrt{2})$, the measure (\ref{ZqYM}) 
yields 
\be
\l[2i\sin\l(\pi{i}{g^2l}n\r)\r]^{\chi(\Sigma)}
=\l[n\r]_q^{\chi(\Sigma)},
\ne
with $q=\exp(-2\pi{g^2}l)$, 
where $[x]_q=\l(q^{x/2}-q^{-x/2}\r)$. 
Na\"ively speaking, 
this is why this theory is called the $q$-deformed theory. 

\section{Five-Dimensional Super Yang-Mills Theory on $\IR^5$}
\label{flatcase}

Let us proceed to a brief explanation about a five-dimensional supersymmetric 
Yang-Mills theory on $\IR^5$, which can be obtained by the dimensional 
reduction in the time direction of six-dimensional maximally supersymmetric 
Yang-Mills theory in a flat Minkowski space. 

In terms of five-dimensional $\N=1$ supermultiplets, 
the vector multiplet in the $\N=2$ theory consists of 
an $\N=1$ vector multiplet and an $\N=1$ hypermultiplet 
in the adjoint representation of the gauge group $G$. 

The vector multiplet consists of a gauge field $v_M$ ($M=1,\cdots,5$), 
a real scalar field $\s$, auxiliary fields $D^{\da}{}_{\db}$, 
and a spinor field $\Psi^{\da}$, 
where the indices $\da$, $\db$ label the components of 
the fundamental representation ${\bf 2}$ of $SU(2)$ $R$-symmetry\footnote{
Although the $R$-symmetry of the $\N=2$ theory is $SO(5)$, 
its subgroup $SO(4)\simeq{SU(2)}\times{SU(2)}$ is manifest in terms of 
the $\N=1$ supermultiplets, 
and we will respect only one of the two $SU(2)$s, in this paper.}. 
The spinor field $\Psi^{\da}$ obeys the symplectic Majorana condition
\be
(\Psi^{\dot{\beta}})^T C_5\epsilon_{\dot{\beta}\dot{\alpha}}
=\l({\Psi}{}_{\dot{\alpha}}\r)^{\dag}
\equiv\bar{\Psi}{}_{\dot{\alpha}},
\ne
where $T$ denotes the transpose, and $\e_{\da\db}$ is the invariant 
tensor of the $SU(2)$ R-symmetry. 
The auxiliary fields $D^{\da}{}_{\db}$ are 
anti-Hermitian and in the adjoint representation of the $SU(2)$ R-symmetry;
\be
D^{\da}{}_{\db}=-(D^{\db}{}_{\da})^\dag,
\qquad
D^{\da}{}_{\dot\g}\,\e^{\dot\g\db}=D^{\db}{}_{\dot\g}\,\e^{\dot\g\da},
\qquad
D^{\da}{}_{\da}=0. 
\ne
Our notations for the charge conjugation 
matrix $C_5$ and the gamma matrices $\G^M$ are explained 
in appendix \ref{notations}. 
Since all the fields are in the adjoint representation of the gauge group $G$, 
they are denoted in the matrix notation as 
\be
\Phi=\Phi^AT^A
\ne
with the normalization\footnote{The gauge group $G$ is assumed to be simple.} 
$\tr{T^AT^B}=\dl^{AB}$. 

On a flat Euclidean space $\IR^5$, the Lagrangian $\L^{(0)}{}_V$ of 
the vector multiplet is given\footnote{The sign of the Lagrangian 
$\L^{(0)}{}_V$ is opposite to the one in the previous 
paper \cite{flatvector}.} by 
\begin{equation}
\tr{
-\frac{1}{4}v_{MN}v^{MN}+\frac{1}{2}D_M \s D^M \s 
+i\bar{\Psi}{}_{\dot{\alpha}}\G^M D_M\Psi^{\dot{\alpha}}
-g\bar{\Psi}{}_{\dot{\alpha}}\l[\s,\,\Psi^{\dot{\alpha}}\r]
-\frac{1}{4}D^{\dot{\alpha}}{}_{\dot{\beta}}D^{\dot{\beta}}{}_{\dot{\alpha}} 
},
\label{flatLv}
\end{equation}
where $v_{MN}$ is the field strength 
\be
v_{MN}=\d_Mv_N-\d_Nv_M+ig\l[v_M,\,v_N\r],
\ne
of the gauge field $v_M$, and the covariant derivatives 
$D_M\Phi$ is given by
\be
D_M\Phi=\d_M\Phi+ig\l[v_M,\,\Phi\r].
\ne

The Lagrangian $\L^{(0)}{}_V$ is left invariant 
under a supersymmetry transformation
\be
&&\delta^{(0)}_{\Sigma}v_M=-i\bar{\Sigma}_{\dot{\alpha}}\G_M\Psi^{\dot{\alpha}},
\qquad
\delta^{(0)}_{\Sigma}\sigma=i\bar{\Sigma}_{\dot{\alpha}}\Psi^{\dot{\alpha}}, 
\nn\\
&&\delta^{(0)}_{\Sigma}\Psi^{\dot{\alpha}}
=-\hf\l(\hf{}v_{MN}\G^{MN}\Sigma^{\dot{\alpha}}
+\G^MD_M\s\Sigma^{\dot{\alpha}}+D^{\da}{}_{\db}\Sigma^{\db}\r),
\label{flatSUSYV}\\
&&\delta^{(0)}_{\Sigma}D^{\dot{\alpha}}{}_{\dot{\beta}}
=i\l[D_M\bar{\Psi}{}_{\dot{\beta}}\G^M\Sigma^{\dot{\alpha}}
+\bar{\Sigma}_{\dot{\beta}}\G^M D_M\Psi^{\dot{\alpha}}
+ig\l(\l[\s,\,\bar{\Psi}{}_{\dot{\beta}}\r]\Sigma^{\dot{\alpha}}
+\bar{\Sigma}_{\dot{\beta}}\l[\s,\,\Psi^{\dot{\alpha}}\r]\r)\r], 
\ne
where the transformation parameter $\Sigma^{\da}$ is also a symplectic 
Majorana spinor;
\be
\bar{\Sigma}{}_{\dot{\alpha}}=
(\Sigma^{\dot{\beta}})^T C_5\epsilon_{\dot{\beta}\dot{\alpha}}.
\ne

The hypermultiplet consists of complex scalar fields $H_{\da}$, 
a spinor field $\Xi$, and auxiliary fields ${F_{H}}_{\a}$ ($\a=1,2$), 
where the index $\a$ is distinct from the one $\da$ of the $SU(2)_R$ symmetry. 
Since they all transform in the adjoint representation under 
a gauge transformation, they are also denoted in the matrix notation. 

They have the free Lagrangian $\L^{(0)}{}_H$ and the interaction 
$\L^{(0)}{}_{\rm int}$ with the vector multiplet, and they are given by
\be
&&\L^{(0)}{}_H=\tr{-D^M\bar{H}^{\da}D_MH_{\da}-i\bar\Xi\G^MD_M\Xi
+\bar{F_H}^{\a}{F_H}_{\a}},
\label{flatLH}\\
&&\L^{(0)}{}_{\rm int}={\rm tr}\bigg[
g^2\l[\s,\,\bar{H}^{\da}\r]\l[\s,\,{H}_{\da}\r]
+igD^{\da}{}_{\db}\l[\bar{H}^{\db},\,H_{\da}\r]-g\bar\Xi\l[\s,\,\Xi\r]
\label{Lint}\\
&&\hskip5cm
-2g\bar\Xi\l[H_{\da},\,\Psi^{\da}\r]+2g\l[\bar{H}^{\da},\,\bar\Psi_{\da}\r]\Xi
\bigg],
\ne
where $\bar{H}^{\da}$ is the complex conjugate of $H_{\da}$, and 
$\bar\Xi=\Xi^{\dag}$. They are left invariant under a supersymmetry 
transformation \cite{HosomichiTerashima} 
\be
&&\dl^{(0)}_{\Sigma}H_{\da}=-i\bar\Sigma_{\da}\Xi,
\nn\\
&&\dl^{(0)}_{\Sigma}\Xi=\l(\G^MD_MH_{\da}+ig\l[\s,\,H_{\da}\r]\r)\Sigma^{\da}
+{F_H}_{\a}\check\Sigma^{\a},
\label{flatSUSYH}\\
&&\dl^{(0)}_{\Sigma}F_H{}_{\a}=i\bar{\check\Sigma}_{\a}
\l[\G^MD_M\Xi-ig\l[\s,\,\Xi\r]-2ig\l[H_{\db},\,\Psi^{\db}\r]\r],
\ne
if accompanied by the transformation (\ref{flatSUSYV}). 
The transformation parameters $\check\Sigma^{\a}$ are 
linearly independent spinors of $\Sigma^{\da}$ and also obey 
the symplectic-Majorana condition
\be
\bar{\check\Sigma}_{\a}=\l(\check\Sigma^{\b}\r)^TC_5\epsilon_{\b\a}. 
\ne

The supersymmetry transformation (\ref{flatSUSYV}),(\ref{flatSUSYH}) yields 
a closed algebra on any field $\Phi$ in the vector multiplet and 
the hypermultiplet 
for the supersymmetry parameters $\Theta^{\da}$, $\Sigma^{\da}$, 
specified in section \ref{SUSYParameter} as
\be
\l[\dl^{(0)}_\Theta,\,\dl^{(0)}_\Sigma\r]\Phi
=\xi^M\d_M\Phi+\dl_G\Phi
\equiv\dl^{(0)}\Phi,
\ne
with 
\be
\xi^M=i\bar\Theta_{\dot\gamma}\G^M\Sigma^{\dot\gamma}, 
\ne
where $\dl_G$ is a gauge transformation with the parameter
\be
\omega=i\l[\bar\Theta_{\dot\gamma}\G^M\Sigma^{\dot\gamma}v_M
+\bar\Theta_{\dot\gamma}\Sigma^{\dot\gamma}\s\r],
\label{gaugeparameter}
\ee
and therefore, on an adjoint field $\Phi$, 
\be
\dl_G\Phi=ig\l[\omega,\,\Phi\r].
\ne

\section{SUSY Parameters on $S^3\times\Sigma$}
\label{SUSYParameter}

In the previous paper \cite{flatvector}, we have considered the 
$\N=1$ supersymmetric Yang-Mills theory on $S^3\times\IR^2$, 
where we have picked up the Killing 
spinor $\e$ on the unit round $S^3$ obeying the Killing spinor equation 
\be
~~~~~~~~\cd_m\e={i\over2}\g_m\e,
\qquad (m=1,2,3)
\label{KSE}
\ee
with the spin connection $\om_m^{ab}$ ($a,b=1,2,3$) 
of the unit round $S^3$ in the covariant derivative
\be
\cd_m\e=\d_m\e+{1\over4}\om_m^{ab}\g^{ab}\e, 
\ne
and have formed the supersymmetry transformation parameter $\Sigma^{\da}$ 
($\da=1,2$) given by 
\be
\Sigma^{\da=1}=\e\otimes\zeta_+, 
\qquad
\Sigma^{\da=2}=C_3^{-1}\e^*\otimes\zeta_-,
\label{SUSYSigma}
\ee
where $*$ denotes the complex conjugation, and 
two-dimensional Weyl spinors\footnote{They were denoted as $\chi_{\pm}$ 
in the previous paper \cite{flatvector}.} 
\be
\zeta_{\pm}={1\over\sqrt{2}}\left(\begin{array}{c}
1\\{\pm{i}}\end{array}\right),
\ne
on $\IR^2$ are the eigenvectors of $i\G^4\G^5$; 
$i\G^4\G^5\zeta_{\pm}=\pm\zeta_{\pm}$.

One of important properties of the parameter $\Sigma^{\da}$ is that 
it obeys the condition
\be
\G^{45}\Sigma^{\da}=-2iN^{\da}{}_{\db}\Sigma^{\db}
\equiv-2\tilde\Sigma^{\da},,
\ne
where the matrix $N^{\da}{}_{\db}$ is defined by 
\be
\l(N^{\da}{}_{\db}\r)=\hf{\s}_3,
\ne
and another condition 
\be
\cd_m\Sigma^{\da}=iN^{\da}{}_{\db}\G_m\Sigma^{\db}
=\G_m\tilde\Sigma^{\db},
\ne
is a direct consequence of the Killing spinor equation (\ref{KSE}) of $\e$. 

The linearly independent parameter $\check\Sigma^{\a}$ is then given by
\be
\check\Sigma^{\a=1}=\e\otimes\zeta_-,
\qquad
\check\Sigma^{\a=2}=C_3^{-1}\e^*\otimes\zeta_+,
\label{SUSYcheckSigma}
\ee
and obeys the similar conditions
\be
\G^{45}\check\Sigma^{\a}=2iN^{\a}{}_{\b}\check\Sigma^{\b}
\equiv-2{\tilde{\check\Sigma}}{}^{\a},
\qquad
\cd_m\check\Sigma^{\a}=-iN^{\a}{}_{\b}\G_m\check\Sigma^{\b}
=\G_m\tilde{\check{\Sigma}}{}^{\a},
\ne
where 
\be
\l(N^{\a}{}_{\b}\r)=\hf{\s}_3. 
\ne

When we replace $\IR^2$ in $S^3\times\IR^2$ by a Riemann surface $\Sigma$, 
we would like to keep using $\Sigma^{\da}$ as 
a supersymmetry transformation parameter on $S^3\times\Sigma$. 
To this end, one needs to introduce a background gauge field by gauging 
the $SU(2)_R$ symmetry. Suppose that the covariant derivative on a Weyl spinor 
$\xi_{+}$ of positive chirality on the Riemann surface $\Sigma$ is given by
\be
\cd_z\xi_{+}=\l(\d_z-{i\over2}\om_z\r)\xi_+,
\ne
with the spin connection $\om_z$ on $\Sigma$, and then 
the gauging of the $SU(2)_R$ symmetry yields the covariant derivative 
\be
\cd_z\Sigma^{\da}=\d_z\Sigma^{\da}+\hf\om_z\G^{45}\Sigma^{\da}
+iA_z^{\da}{}_{\db}\Sigma^{\db}
=\d_z\Sigma^{\da}+i\l(A_z^{\da}{}_{\db}-\om_z{N}^{\da}{}_{\db}\r)\Sigma^{\db}
\label{covSigma}
\ee
on the parameter $\Sigma^{\da}$. We here have identified 
the local complex coordinates $z,\bz$ on $\Sigma$ with 
$z=x^4+ix^5$, $\bz=x^4-ix^5$. 
If one takes the background gauge field 
$A_z^{\da}{}_{\db}$ as 
\be
A_z^{\da}{}_{\db}=\om_z{N}^{\da}{}_{\db},
\ne
it is obvious from (\ref{covSigma}) that 
the parameter $\Sigma^{\da}$ in (\ref{SUSYSigma}) yields 
a covariantly constant spinor on $\Sigma$ so that $\Sigma^{\da}$ 
is well-defined on the Riemann surface $\Sigma$. 

As for the other parameter $\check\Sigma^{\a}$, we will introduce 
a background gauge field $\check{A}_z^{\a}{}_{\b}=-\om_z{N}^{\a}{}_{\b}$ 
to make $\check\Sigma^{\a}$ covariantly constant and so well-defined 
on $\Sigma$.

This ``partial twisting'' \cite{Bershadsky,Wallcrossing} 
affects the spin representations of the fields 
carrying the indices of the $SU(2)_R$ symmetry or the index $\a$ of 
another $SU(2)$ symmetry. 
In order to see this and also for later use, 
it is convenient to give the gauge field $v^M$ and 
the spinor $\Psi^{\da}$ in the vector multiplet 
in terms of three-dimensional tensors and spinors as
\be
&&v^{m}\quad (m=1,2,3),
\qquad
v_z=\hf\l(v_4-iv_5\r),
\qquad
v_{\bz}=\hf\l(v_4+iv_5\r),
\nn\\
&&\Psi^{\da=1}=\lambda\otimes\zeta_++
\psi\otimes\zeta_-,
\qquad
\Psi^{\dot{\alpha}=2}=C^{-1}_3 \psi^{*}\otimes\zeta_{+}
+C^{-1}_3 \lambda^{*}\otimes\zeta_{-},
\nn\\
&&D=D^{1}{}_{1}+g^{\bzz}\,v_{\bzz},
\qquad
F_z=\hf\,D^{1}{}_{2}, 
\qquad
\bar{F}_{\bz}=\hf\,D^{2}{}_{1}. 
\ne
As for the hypermultiplet, 
\be
\l(H_{\da}\r)=\left(\begin{array}{c}
H_1\\{H_2}\end{array}\right)=\left(\begin{array}{c}
\tilde{H}\\{\l(H\r)^*}\end{array}\right),
\nn\\
\Xi=\tilde\chi\otimes\zeta_++C_3^{-1}\l(\chi\r)^*\otimes\zeta_-.
\ne

In terms of these fields, after the partial twisting, 
in the vector multiplet,
$\la$ becomes a scalar on $\Sigma$, while $\psi$ becomes a $(1,0)$-form 
$\psi_z$. 
The auxiliary fields $D$ and $F_z$ are a scalar and a $(1,0)$-form, 
respectively. 
In the hypermultiplet, the scalars $\tilde{H}$, ${H}$ become Weyl spinors 
of positive chirality, while $\tilde\chi$, $\chi$ are unaffected to remain 
Weyl spinors of positive chirality. The auxiliary fields ${F_H}_{1}$ and 
${F_H}_{2}$ become Weyl spinors of negative chirality and positive chirality, 
respectively.

\hspace{10cm}
\section{Supersymmetry on $S^3\times\Sigma$}
\label{SUSYalg}

When going onto the $S^3\times\Sigma$, 
we turn on the spin connections of $S^3\times\Sigma$ and 
the background gauge fields $A^{\da}{}_{\db}$, $A^{\a}{}_{\b}$ 
in the covariant derivatives in the supersymmetry transformation 
(\ref{flatSUSYV}), (\ref{flatSUSYH}). However, 
the supersymmetry transformation (\ref{flatSUSYV}), (\ref{flatSUSYH})
no longer yields a closed algebra, since the supersymmetry parameters 
$\Sigma^{\da}$, $\check\Sigma^{\a}$ aren't covariantly constant along 
the $S^3$, although the gauging of the $SU(2)_R$ symmetry and 
the other $SU(2)$ symmetry 
makes them covariantly constant along the Riemann surface $\Sigma$.

In order to give a closed algebra of the supersymmetry transformations 
on $S^3\times\Sigma$, one needs to modify the transformations 
by adding the terms
\be
\dl'_{\Sigma}D^{\da}{}_{\db}=-2i\l(\bar{\tilde{\Sigma}}_{\db}\Psi^{\da}
+\bar\Psi_{\db}\tilde{\Sigma}^{\da}\r),
\ne
to (\ref{flatSUSYV}) for the vector multiplet and 
\be
\dl'_{\Sigma}\Xi=2H_{\da}\tilde{\Sigma}^{\da}, 
\qquad
\dl'_{\Sigma}{F_H}{}_{\a}={i\over2}\bar{\check{\Sigma}}_{\a}\G^{45}\Xi,
\ne
to  (\ref{flatSUSYH}) for the hypermultiplet.

The modified transformation $\dl_\Sigma=\dl^{(0)}_\Sigma+\dl'_\Sigma$ 
indeed gives the closed supersymmetry algebra on any field $\Phi$ 
in the vector multiplet and the hypermultiplet as
\be
\l[\dl_\Theta,\,\dl_\Sigma\r]\Phi={\pounds}_\xi\Phi+\dl_G\Phi+\dl_R\Phi,
\ne
with the parameters (\ref{SUSYSigma}),(\ref{SUSYcheckSigma}), and 
their analogues $\Theta^{\da}$, $\check\Theta^{\a}$ where 
$\e$ is replaced by a solution $\eta$ to (\ref{KSE}) 
in (\ref{SUSYSigma}),(\ref{SUSYcheckSigma}), 
respectively. 
The transformation 
$\dl_G$ is the same gauge transformation as before with the parameter 
(\ref{gaugeparameter}), and $\dl_R$ is the transformation of a $U(1)$ 
subgroup of the $SU(2)_R$ symmetry and the other $SU(2)$ symmetry and 
is given by 
\be
\dl_R\Phi^{\da}=2i\l(\bar{\tilde{\Theta}}_{\db}\Sigma^{\da}
-\bar{\tilde{\Sigma}}_{\db}\Theta^{\da}\r)\Phi^{\db},
\qquad
\dl_R\Phi_{\a}=-2i\Phi_{\b}
\l(\bar{\check{\Theta}}_{\a}\tilde{\check{\Sigma}}^{\b}
-\bar{\check{\Sigma}}_{\a}\tilde{\check{\Theta}}^{\b}
\r),
\ne
for the fundamental representation $\Phi^{\da}$ of the $SU(2)_R$ and 
the anti-fundamental representation $\Phi_{\a}$ of the other $SU(2)$. 
The partial derivative which was in the translation is replaced by 
the Lie derivative $\pounds_\xi$ with respect to the vector 
$\xi^M=i\bar\Theta_{\dot\gamma}\G^M\Sigma^{\dot\gamma}$. 
Here we have defined the Lie derivative $\pounds_\xi$ on a five-dimensional 
spinor field $\Psi$ as
\be
\pounds_\xi\Psi=\xi^M\cd_M\Phi+{1\over4}\l(\cd_M\xi_N\r)\G^{MN}\Psi.
\ne
Note that the translation in $\IR^5$ is extended to 
the infinitesimal diffeomorphism with the parameter $\xi^M$ on the curved space 
$S^3\times\Sigma$, and the diffeomorphism also transforms the background 
vielbein non-trivially. Therefore, the Lorentz transformation in the second 
term of the above definition is needed to compensate the diffeomorphism 
in order to keep the background vielbein intact. 

In terms of the three-dimensional fields, the supersymmetry 
transformation yields 
\be
&&\dl_\Sigma v_m=-i\l[\bar{\e}\g_m\la-\bar{\la}\g_m\e\r],
\qquad
\dl_\Sigma v_z=-\bar{\e}\psi_z, 
\qquad
\dl_\Sigma\s =i\l[\bar{\e}\la-\bar{\la}\e\r],
\nn\\
&&\dl_\Sigma\la=-\hf\l[\hf\,v_{mn}\g^{mn}
+\g^mD_m\sigma+D\r]\e,
\nn\\
&&\dl_\Sigma\psi_z=\l[iv_{mz}\g^m\e-iD_z\s\e-F_z\,C^{-1}_3\e^{*}\r],
\label{SUSYV}\\
&&\dl_\Sigma D=i\bigg[D_m \bar{\la}\g^m\e+\bar{\e}\g^mD_m\la
+ig\l(\l[\s,\,\bar{\lambda}\r]\epsilon+\bar{\e}\l[\s,\,\la\r]\r)
+{i\over2}\l(\bar{\e}\la -\bar{\la}\e\r)
\bigg],
\nn\\
&&\delta_\Sigma F_z =i\l[-\epsilon^T C_3\g^m D_m\psi_z
+2i\epsilon^TC_3D_z\lambda+ig\epsilon^TC_3\l[\s,\,\psi_z\r]
-{i\over2}\e^T C_3 \psi_z
\r],
\ne
for the vector multiplet, and
\be
&&\dl_\Sigma\tilde{H}=-i\bar\e\tilde\chi,
\qquad
\dl_\Sigma{H}=-i\bar\e\chi,
\nn\\
&&\dl_\Sigma\tilde\chi=\l[D_m\tilde{H}\g^m+ig\l[\s,\,\tilde{H}\r]
+i\tilde{H}\r]\e+\l[-2i\l(D_zH\r)^*+F_H{}_2\r]C_3^{-1}\e^*,
\nn\\
&&\dl_\Sigma\chi=\l[D_m{H}\g^m+ig\l[\s,\,{H}\r]
+i{H}\r]\e+\l[2i\l(D_z\tilde{H}\r)^*-\l(F_H{}_1\r)^*\r]C_3^{-1}\e^*,
\nn\\
&&\dl_\Sigma{F_H}{}_{1}^*=i\bigg[
\l(D_m\chi\r)^TC_3\g^m+2i\l(D_z\tilde\chi\r)^\dag+{i\over2}\chi^TC_3
+ig\l[\s,\,\chi^T\r]C_3
\label{SUSYH}\\
&&\hskip6cm
+2ig\l(\l[\tilde{H}^*,\,\bar\psi\r]
+\l[H,\,\la^T\r]C_3\r)
\bigg]\e,
\nn\\
&&\dl_\Sigma{F_H}{}_{2}^*=-i\bigg[
\l(D_m\tilde\chi\r)^{\dag}\g^m+2i\l(D_z\chi\r)^TC_3+{i\over2}{\tilde\chi}^\dag
-ig\l[\s,\,{\tilde\chi}^\dag\r]
\nn\\
&&\hskip5.5cm
-2ig\l(\l[\tilde{H}^*,\,\bar\la\r]
+\l[H,\,\psi^T\r]C_3\r)
\bigg]C_3^{-1}\e^*,
\ne
for the hypermultiplet. 

The Lagrangians $\L^{(0)}_V$, $\L^{(0)}_H$, $\L^{(0)}_{\rm int}$ 
in the $\IR^5$ need to be covariantized in order to put them 
on the curved space $S^3\times{\Sigma}$. In addition, the gauging 
of $SU(2)_R$ symmetry and the other $SU(2)$ symmetry in them will be 
done to consider the supersymmetry transformation 
(\ref{SUSYV}), (\ref{SUSYH}) 
with the parameters (\ref{SUSYSigma}), (\ref{SUSYcheckSigma}) 
of the Lagrangians. 

However, these aren't enough to obtain 
invariant Lagrangians under the transformation (\ref{SUSYV}), (\ref{SUSYH}), 
and to this end, one needs additional terms to the covariantized Lagrangians. 
In fact, it turns out that 
the additional terms to $\L^{(0)}_V$ of the vector multiplet are given by
\be
\L'_V&=&\tr{
N^{\da}{}_{\db}\bar\Psi_{\da}\Psi^{\da}+
\l(iN^{\da}{}_{\db}D^{\db}{}_{\da}-\hf\ve^{ij}v_{ij}\r)\s+\s^2
+\hf\om_{\rm c.s.}
}
\nn\\
&=&\tr{\hf{g}^{\bzz}\bar{\psi}_{\bz}\psi_z+\bar{\la}\la+\s\s
+i\s\l(D-2g^{\bzz}v_{\bzz}\r)+\hf\om_{\rm c.s.}},
\ne
with the Chern-Simons term 
\be
\om_{\rm c.s.}=\ve^{mnk}\l(v_m\d_n{v}_k+{i\over3}g\,v_m\l[v_n,\,v_k\r]\r), 
\ne 
and to $\L^{(0)}_H$ of the hypermultiplet, 
\be
\L'_H=\tr{-{i\over2}\bar\Xi\,\G^{45}\Xi-\bar{H}^{\da}H_{\da}}
=-\tr{\hf\l(\bar{\tilde{\chi}}\tilde\chi+\bar\chi\chi\r)
+\l(\tilde{H}^*\tilde{H}+H^*H\r)}.
\ne
With these terms, one can verify that the Lagrangian 
\be
&&\L_V=\L^{(0)}_V+\L' 
\nn\\
&&=
{\rm tr}\bigg[
-\frac{1}{4}(v_{mn})^2
-g^{\bzz}v_{m\bz}v_{mz}
+\frac{1}{2}(D_m\sigma)^2
+g^{\bzz}D_z\sigma D_{\bar{z}}\sigma 
-\hf{}D^2
+g^{\bzz}v_{\bzz}\,D
\nn\\
&&\hskip1.2cm
-g^{\bzz}\bar{F}_{\bz}F_z
+2i\bar{\la}\g^mD_m\la
-ig^{\bzz}\bar{\psi}_{\bz}\g^mD_m\psi_z
-2g^{\bzz}\bar{\psi}_{\bz}\,D_z\la
-2g^{\bzz}D_{\bz}\bar\la\,\psi_z
\nn\\
&&\hskip1.2cm
-g\l(2\bar{\la}\l[\s,\,\la\r]+g^{\bzz}\bar{\psi}_{\bz}\l[\s,\,\psi_z\r]\r)
\nn\\
&&\hskip1.2cm
+\hf{g}^{\bzz}\bar{\psi}_{\bz}\psi_z+\bar{\la}\la+\s\s
+i\s\l(D-2g^{\bzz}v_{\bzz}\r)
+\hf\om_{\rm c.s.}
\bigg]
\ne
of the vector multiplet is left invariant under 
the supersymmetry transformation (\ref{SUSYV}), and the total Lagrangian 
$\L=\L_V+\L^{(0)}_H+\L^{(0)}_{\rm int}+\L'_H$ under (\ref{SUSYV}), 
(\ref{SUSYH}).

\section{Localization}
\label{Localization}

In this section, we will compute the partition function of 
the $\N=2$ theory on the ${S}^3\times\Sigma$ 
by using the localization method. 
In the next subsection, we will calculate the contribution from 
the vector multiplet to the partition function, which can be 
regarded as the extension of our previous results \cite{flatvector} 
about the $\N=1$ theory on ${S}^3\times\IR^2$ for the ${S}^3\times\Sigma$. 
And we will see that the $\N=1$ theory on the ${S}^3\times\Sigma$ yields 
the partition function of two-dimensional Yang-Mills theory on $\Sigma$.

In subsection \ref{LocalizationH}, we will proceed to the calculations 
of the contribution from the hypermultiplet and will find that 
it yields no contributions to the partition function of the 
$\N=2$ theory on the ${S}^3\times\Sigma$. 

\subsection{The Contribution from the Vector Multiplet}
\label{LocalizationV}

In the Lagrangian $\L_V$, the kinetic terms of the bosonic fields 
$\s$, $D$, $F_z$, and $\bar{F}_{\bz}$ have the wrong sign. 
In order to make the path-integral well-defined, 
they need to be analytically continued. 
Therefore, we will regard the auxiliary field $D$ as a real field and 
will replace the scalar field $\s$ by $i\s$, 
where the latter $\s$ takes the real value\footnote{
In the previous paper \cite{flatvector}, after the analytic continuation, 
$\s$ was regarded as taking pure imaginary values.}. 
In addition, $\bar{F}_\bz=\l(F_z\r)^*$.

To carry out the localization procedure, we will define the BRST transformation 
by setting $\bar\e$ to zero in the supersymmetric transformation (\ref{SUSYV}) 
and by replacing the Grassmann odd parameter $\e$ by a Grassmann even one.
It yields 
\be
&&\dl_Q v_m=-i\bar{\la}\g_m\e,
\qquad
\dl_Q v_z=0, 
\quad
\dl_Q v_{\bz}=\bar\psi_z\e, 
\qquad
\dl_Q\s =\bar{\la}\e,
\nn\\
&&\dl_Q\la=-\hf\l[\hf\,v_{mn}\g^{mn}+i\g^mD_m\sigma+D\r]\e,
\qquad
\dl_Q\bar\la=0,
\nn\\
&&\dl_Q\psi_z=\l[iv_{mz}\g^m+D_z\s\r]\e,
\qquad
\dl_Q\bar\psi_\bz=\bar{F}_\bz\,\e^TC_3,
\label{BRSTV}\\
&&\dl_Q D=\bigg[-iD_m \bar{\la}\g^m\e
+ig\l[\s,\,\bar{\lambda}\r]\e-{1\over2}\bar{\la}\e
\bigg],
\nn\\
&&\delta_Q F_z =\e^TC_3\l[-i\g^m D_m\psi_z
-2D_z\lambda-ig\l[\s,\,\psi_z\r]
+{1\over2}\psi_z\r],
\quad
\delta_Q \bar{F}_\bz =0,
\ne
which is in fact nilpotent; $\dl_Q^2=0$. 
Using the BRST transformation (\ref{BRSTV}), we will modify the Lagrangian 
$\L_V$ into $\L_V-{t}\,\L_{VQ}$ with a parameter $t$, where 
\be
\L_{VQ}
=\dl_Q\tr{
\l(\dl_Q\la\r)^\dag\la
+\hf{g}^{\bzz}\l(\dl_Q\psi_z\r)^\dag\psi_z
+\hf{g}^{\bzz}\bar{\psi}_{\bz}\l(\dl_Q\bar\psi_{\bz}\r)^\dag}.
\label{LVQ}
\ee
The bosonic part of the extra Lagrangian $\L_{VQ}$ gives 
\be
&&\L_{VQ}^{(B)}
=\hf{\rm tr}\bigg[
{1\over4}v^{mn}v_{mn}
+g^{\bzz}g^{mn}v_{m\bz}v_{nz}
+\hf{D}^m\s{D}_m\s
+g^{\bzz}D_{\bz}\s{}D_z\s 
+\hf\,D^2
\nn\\
&&\hskip2.7cm
+g^{\bzz}\bar{F}_\bz{}F_z
+ik^m{g}^{\bzz}\l(v_{mz}D_{\bz}\s-v_{m\bz}D_z\s\r)
+ik_m\e^{mnk}{g}^{\bzz}v_{n\bz}v_{kz}
\bigg]
\ne
where the Killing vector $k_m$ was defined by
\be
k_m=\bar\e\g_m\e
\ne
with the normalization $\l(\bar\e\e\r)=1$.
On the other hand, the fermionic part of $\L_{VQ}$ gives 
\be
&&\L_{VQ}^{(F)}=
i{\rm tr}\bigg[
-\bar\la\g^m{D}_m\la-{i\over2}\bar\la\la+g\bar\la\l[\s,\,\la\r]
+\hf{g}^{\bzz}\bar{\psi}_{\bz}\g^m{D}_m\psi_z
-{i\over4}g^{\bzz}\bar{\psi}_{\bz}\psi_z
\nn\\
&&\hskip2.3cm
+\hf{g}^{\bzz}\l({g}\bar{\psi}_\bz\l[\s,\,\psi_z\r]
-ik_m\bar{\psi}_\bz\g^m\psi_z+2i\bar\la{D}_{\bz}\psi_z
-i\bar{\psi}_{\bz}{D}_z\la+ik_m\bar{\psi}_\bz\g^mD_z\la\r)
\bigg]. 
\ne

In the large $t$ limit, $t\to\infty$, the fixed point, 
which is a solution to 
\be
\l[\hf\,v_{mn}\g^{mn}+i\g^mD_m\sigma
+D\r]\e=0,
\qquad
\l[v_{mz}\g^m-iD_z\s\r]\e=0,
\qquad
F_z=0,
\ne
gives
the dominant contribution to the partition function. 
In fact, the fixed point is given \cite{Kapustin} by
\be
&&v_m=0, 
\quad
D=0,
\quad
F=0,
\quad
v_z=v_z(z,\bz), 
\quad
\s=\s(z,\bz), 
\quad
D_z\s=0.
\label{fixedpt}
\ee
Substituting the background (\ref{fixedpt}) into the Lagrangian $\L_V$, 
one finds that the additional Lagrangian $\L'{}_V$ only contributes and yields 
\be
\L_{YM}=\tr{-\s\s+2\s\,{g}^{\bzz}v_{\bzz}},
\label{YM}
\ee
which would give the action of the two-dimensional Yang-Mills theory 
if one could integrate out the scalar field $\s$. However, in this case, 
the scalar field $\s$ must obey the condition $D_z\s=0$, and 
it isn't allowed to perform the Gaussian integration over the whole 
functional space of the $\s$. 

Around the fixed points, we will perform the path integral 
over the quantum fluctuations. Since the bosonic fields $\s$, $v_z$, 
and $v_{\bz}$ have a non-trivial background as the fixed point, 
we will expand the fields as
\be
\s=\s(z,\bz)+{1\over\sqrt{t}}\t{\s}(x^m,z,\bz), 
\qquad
v_z=v_z(z,\bz)+{1\over\sqrt{t}}\t{v}_z(x^m,z,\bz), 
\ne
while the other fields are rescaled as 
$\Phi\to(1/\sqrt{t})\tilde\Phi$.

One also needs the gauge-fixing procedure for the computation of 
the path integral. We will follow \cite{Kapustin} and add 
to $\L_{VQ}$ the gauge-fixing term and the ghost term 
\be
\tr{\bar{C}\cd_m{D}^mC+B\cd^mv_m}. 
\label{gaugefixing}
\ee

There remains the residual gauge symmetry, under which 
\be
\s\to\s+ig\l[\om(z,\bz),\s\r], 
\qquad
v_z\to{v}_z-D_z\om(z,\bz),
\label{residualgauge}
\ee
where the gauge transformation parameter $\om$ is constant along the $S^3$. 
The symmetry is the redundancy of the backgrounds $\s$ and $v_z$, but not 
of the fluctuations, and the gauge fixing procedure can be carried out in 
a similar way to the two-dimensional Yang-Mills theory, as explained in 
\cite{Lecture,qYM}.  
Following \cite{Lecture,qYM}, 
one can make use of the residual symmetry (\ref{residualgauge}) 
to put the background $\s(z,\bz)$ 
in the Cartan subalgebra of the Lie algebra of $G$ such that 
\be
\s(z,\bz)=\sum_{i=1}^{r}\s_i\,{}H_i.
\label{Cartansigma}
\ee
Recall that $H_i$ ($i=1,\cdots,r$) are the generators of 
the Cartan subalgebra of $G$ of rank $r$, 
and the localization condition $D_z\s=0$ in (\ref{fixedpt}) implies that  
the background $v_z(z,\bz)$ should also be in the Cartan subalgebra as 
\be
v_z(z,\bz)=\sum_{i=1}^{r}v^i{}_z(z,\bz){}H_i, 
\label{Cartanvz}
\ee
and furthermore that 
$\s_i$ ($i=1,\cdots,r$) are constant with respect to the local 
coordinates $z,\bz$ of the $\Sigma$. 

Therefore, to impose the gauge-fixing condition (\ref{Cartansigma}), 
we will follow the same BRST quantization procedure 
as for the two-dimensional Yang-Mills theory 
in \cite{Lecture,qYM}. 
Introducing another auxiliary field 
\be
b(z,\bz)=\sum_{\a\in\La}b^{\a}(z,\bz)E_{\a},
\ne
and another pair of ghost fields
\be
c(z,\bz)=\sum_{\a\in\La}c^{\a}(z,\bz)E_{\a},
\qquad
\bar{c}(z,\bz)=\sum_{\a\in\La}\bar{c}^{\a}(z,\bz)E_{\a},
\ne
where $\La$ is the set of all the root\footnote{
Here, $\a$ denotes a root of the Lie algebra of $G$, but not 
the index of the $SU(2)$ symmetry. We hope which one 
we mean will be clear from the context.} of the Lie algebra of $G$, 
and the root generators $E_{\a}$ satisfy the algebra
\be
\l[H_i,\,E_{\a}\r]=\a_i\,{E}_\a, 
\qquad
\l[E_\a,\,E_{-\a}\r]=\sum_{i=1}^{r}\a_i\,{H}_i\equiv\a\cdot{H},
\ne
we will add another gauge-fixing term and the ghost term 
\be
\sum_{\a\in\La}\l[ib^{-\a}\s^{\a}-ig\as\bar{c}^{-\a}c^{\a}\r],
\ne
where $\as=\sum_{i=1}^{r}\a^i\s^i$. 
After the integration over the root part $\s^\a$, the auxiliary field 
$b^\a$ and the ghosts $\bar{c}^\a$, $c^\a$, 
the path-integral measure of the scalar field $\s(z,\bz)$ results in 
the finite-dimensional integral over $\s_i$ ($i=1,\cdots,r$), 
the ghosts give the one-loop determinant 
\be
\prod_{\a\in\Lambda}\Det{0,0}{ig\as},
\label{DetFP}
\ee 
which is the same contribution as the ghosts do in the two-dimensional 
Yang-Mills theory. 

Now, let us proceed to the calculations of the one-loop determinants 
of the vector multiplet. 
To this end, we will follow the same procedure as in \cite{Kapustin,Hosomichi}, 
- expanding all the fields in terms of the spherical harmonics on $S^3$ and 
performing the Gaussian integration over them. 

On the unit round $S^3$, we will use the vielbein $e^a=e^a{}_mdx^m$ obeying 
$de^a=\e^{abc}e^b\wedge{e}^c$ and the spin connection $\om^{ab}=\e^{abc}e^c$. 
We define the three-dimensional Hodge duality as
\be
*e^a=\hf\e^{abc}e^b\wedge{e}^c, 
\qquad
*\l(e^a\wedge{e}^b\r)=\e^{abc}e^c, 
\ne
with $*1$ the volume form, and two operators $\imath_k$ and $S^a$ 
acting on the vielbein by
\be
\imath_k{e}^a=e^a{}_m{k}^m=k^a, 
\qquad
S^a{e}^b=i\e^{abc}e^c,
\ne
where $k^m=\bar{\e}\g^m\e$ is the Killing vector field. 

Expanding the Lagrangian $\L_{VQ}$ in terms of the fluctuations 
up to quadratic order, one finds that the bosonic part $\L_{VQ}^{(B)}$ 
in a differential form notation gives 
\be
&&\hf{\rm tr}\bigg[
\hf{d\tilde{v}}\wedge*{d\tilde{v}}+\hf{D}\tilde\s\wedge*D\tilde{\s}
+g^{\bzz}\l(d\tilde{v}_\bz-D_\bz\tilde{v}\r)
\wedge*\l(d\tilde{v}_z-D_z\tilde{v}\r)
\label{LVQB}\\
&&\quad
+g^{\bzz}\l(D_\bz\tilde\s-ig\l[\s,\,\tilde{v}_\bz\r]\r)
\l(D_z\tilde\s-ig\l[\s,\,\tilde{v}_z\r]\r)*1
+ig^{\bzz}\l(D_\bz\tilde\s-ig\l[\s,\,\tilde{v}_\bz\r]\r)\imath_k
\l(d\tilde{v}_z-D_z\tilde{v}\r)*1
\nn\\
&&\quad-ig^{\bzz}\l(D_z\tilde\s-ig\l[\s,\,\tilde{v}_z\r]\r)\imath_k
\l(d\tilde{v}_\bz-D_\bz\tilde{v}\r)*1
-g^{\bzz}\l(d\tilde{v}_\bz-D_\bz\tilde{v}\r)
\wedge*\l[(k\cdot{S})\l(d\tilde{v}_z-D_z\tilde{v}\r)\r]
\bigg],
\ne
with $\l(k\cdot{S}\r)=k^aS^a$, where the form notation denotes 
\be
\tilde{v}=\tilde{v}_mdx^m, 
\qquad
D\tilde\s=d\t\s-ig\l[\s,\,\tilde{v}\r]
=\l(\d_m\t\s-ig\l[\s,\,\tilde{v}_m\r]\r)dx^m.
\ne
Note that the gauge fields in the covariant derivatives $D_z$ and $D_\bz$ here 
are the background $v_z$ and $v_\bz$, respectively. 

For brevity, we will omit the tilde $\tilde{~}$ for the fermionic fluctuations, 
and then the fermionic part $\L_{VQ}^{(F)}$ gives 
\be
&&{\rm tr}\bigg[
-i\bar\la\g^m\cd_m\la+\hf\bar\la\la+ig\bar\la\l[\s,\,\la\r]
+{i\over2}g^{\bzz}\bar\psi_{\bz}\g^m\cd_m\psi_z
+{1\over4}g^{\bzz}\bar\psi_{\bz}\psi_z
\label{LVQF}\\
&&\qquad
+{i\over2}gg^{\bzz}\bar\psi_{\bz}\l[\s,\,\psi_z\r]
+\hf{g}^{\bzz}k_m\bar\psi_\bz\g^m\psi_z
+g^{\bzz}D_\bz\bar\la\cdot\psi_z
+\hf{g}^{\bzz}\bar\psi_\bz\l(1-k_m\g^m\r)D_z\la
\bigg],
\ne
with the covariant derivatives $D_z$ and $D_\bz$ including only the background 
gauge fields $v_z$ and $v_\bz$, respectively. 

In terms of the scalar spherical harmonics $\vp_\lmm$ 
($l=0,1/2,1,3/2,\cdots$; $-l\le{m}\leq{l}$; $-l\leq\mt\leq{l}$),
with the properties
\be
&&d^\dag{d}\vp_{\lmm}=-*d*d\vp_{\lmm}=4l(l+1)\vp_{\lmm},
\nn\\
&&\l(\vp_{\lmm}\r)^*=(-)^{m+\mt}\vp_{l,-m,-\mt},
\nn\\
&&\int_{S^3}\, \l(\vp_{l',m',\mt'}\r)^*\,\vp_\lmm\,*1=
\dl_{l'l}\dl_{m,m'}\dl_{\mt,\mt'},
\ne
on the $S^3$, 
the fields $\t\s$, $\t{v}_z$ are expanded as 
\be
\t\s=\sumlmm\t\s_\lmm(z,\bz)\vp_\lmm(x),
\qquad
\t{v}_z=\sumlmm\t{v}_{z,\lmm}(z,\bz)\vp_\lmm(x). 
\ne
The vector spherical harmonics on the $S^3$ 
are combined by the scalar spherical harmonics $\vp_\lmm$ and 
the vielbein $e^a$. In particular, we will take the vielbein 
as the eigenstate of the operators $S^aS^a$ and $S^3$ as
\be
e^{\pm1}=\mp{1\over\sqrt{2}}\l(e^1\pm{i}e^2\r),
\qquad
e^0=e^3, 
\ne
and form the vector spherical harmonics on the $S^3$, 
\be
E_{J,M;l,\mt}=\sum_{m=-l}^{l}\sum_{s=-1}^{+1}
\langle l,m; 1,s | J,M \rangle\rangle \,\vp_\lmm \, e^s,
\ne
where $\langle l,m; 1,s | J,M \rangle\rangle$ are 
the Clebsch-Gordan coefficients of the spin $l$ representation 
and the spin $1$ representation of the $SU(2)$ group into the spin $J$ 
representation, with $J=(l-1),l,(l+1)$. 
The Clebsch-Gordan coefficients are listed in appendix \ref{ClebschGordan}.
They have the properties
\be
&&*d\,E_{l+1,M;l,\mt}=2(l+1)E_{l+1,M;l,\mt},
\qquad
d*\,E_{l+1,M;l,\mt}=0,
\qquad
(l=0,\hf,1,\cdots)
\nn\\
&&*d\,E_{l-1,M;l,\mt}=-2l\,E_{l-1,M;l,\mt},
\qquad
d*\,E_{l-1,M;l,\mt}=0,
\qquad
(l=1,{3\over2},2,\cdots)
\nn\\
&&*d\,E_{l,M;l,\mt}=0,
\qquad
E_{l,M;l,\mt}=-{i\over2}\sqrt{1\over{l(l+1)}}d\vp_\lmm,
\qquad
(l=\hf,1,{3\over2},\cdots)
\ne
and form the orthonormal basis
\be
\int_{S^3}\, \l(E_{J',M';l',\mt'}\r)^*\wedge\,*E_{J,M;l,\mt}
=\dl_{J,J'}\dl_{M,M'}\dl_{l,l'}\dl_{m,m'}.
\ne
In terms of them, the gauge field $\t{v}_m$ is expanded as 
$\t{v}=\t{v}_{+}+\t{v}_{-}+\t{v}_{L}$, where
\be
&&\t{v}_{+}
=\sum_{l=0}^{\infty}
\sum_{M=-l-1}^{l+1}\sum_{\mt=-l}^{l}
\t{v}_{l+1,M;l,\mt}(z,\bz)E_{l+1,M;l\mt}(x),
\nn\\
&&\t{v}_{-}
=\sum_{l=1}^{\infty}
\sum_{M=-(l-1)}^{l-1}\sum_{\mt=-l}^{l}
\t{v}_{l-1,M;l,\mt}(z,\bz)E_{l-1,M;l\mt}(x),
\nn\\
&&\t{v}_{L}
=\sum_{l=1/2}^{\infty}
\sum_{M=-l}^{l}\sum_{\mt=-l}^{l}
\t{v}_{l,M;l,\mt}(z,\bz)E_{l,M;l\mt}(x)
\nn\\
&&\hskip5mm
=-{i\over2}d
\sum_{l=1/2}^{\infty}
\sum_{M=-l}^{l}\sum_{\mt=-l}^{l}\sqrt{1\over{l(l+1)}}
\t{v}_{l,M;l,\mt}(z,\bz)E_{l,M;l\mt}(x)=d\t{u}_L,
\ne
with $\t{v}_{\pm}$ the transverse modes and $\t{v}_{L}$ the longitudinal mode. 

Substituting these expansions into the bosonic part (\ref{LVQB}), 
one sees that the longitudinal mode $\t{v}_{L}$ can be eliminated in 
$\L_{VQ}^{(B)}$ 
by shifting the fields $\t{v}_z$ and $\t\s$ as 
\be
\t{v}_z \to \t{v}_z+D_z\t{u}_L,
\qquad
\t\s \to \t\s +ig\l[\s,\,\t{u}_L\r]. 
\ne
Therefore, the longitudinal mode $\t{v}_{L}$ appears only in the 
gauge fixing term (\ref{gaugefixing}), which up to quadratic order 
yields 
\be
\bar{C}d*dC+Bd*\t{v}=\bar{C}d*dC+Bd*d\t{u}_L. 
\ne
It is obvious that the one-loop determinant from $B$ and $\t{u}_L$ 
exactly cancels the one-loop determinant from the ghosts $\bar{C}$ and $C$. 

As for the operators $\imath_k$ and $(k\cdot{S})$ with the Killing 
vector $k_a$ appearing in (\ref{LVQB}), one will take the Killing 
spinor $\e$ as constant, and then the Killing vector $k^a=\bar\e\g^a\e$ 
is also constant. Since $k^ak_a=1$ with the normalization $\bar\e\e=1$, 
we will choose it as $k^a=\dl^{a}{}_{3}$, as in \cite{Kapustin}. 
Therefore, one obtains the formulas
\be
&&\int_{S^3}\,\l(\vp_{l',m',\mt'}\r)^*\imath_k{E}_{J,M;l,\mt}*1
=\int_{S^3}\,\l(\vp_{l',m',\mt'}\r)^*\,e^3\wedge*E_{J,M;l,\mt}
=\dl_{l,l'}\dl_{\mt,\mt'}
\langle l,m'; 1,s=0 | J, M \rangle\rangle,
\nn\\
&&\int_{S^3}\,\l({E}_{J',M';l',\mt'}\r)
\wedge*\l[\l(k\cdot{S}\r){E}_{J,M;l,\mt}\r]
=\dl_{l,l'}\dl_{\mt,\mt'}\,\langle\langle J',M' | S^3 | J, M \rangle\rangle.
\ne
For the coefficients $\langle l,m'; 1,s=0 | J, M \rangle\rangle$, 
$\langle\langle J',M' | S^3 | J, M \rangle\rangle$, see the list 
in appendix \ref{ClebschGordan}. 

We are not interested in the overall constant of the partition function, 
but in its dependence on the background $\s^i$ and $v^i_z$. 
The Cartan part $\t\Phi^i$ of the fluctuations 
and the root part $\t\Phi^\a$ of them are completely decoupled 
from each other, and the Cartan part doesn't yield the contributions 
which has the dependence of the background. We will thus focus on the 
contributions from the root part of the fluctuations, but one can 
easily see that the contributions from the Cartan part of the fluctuations 
can be obtained by setting $\as$ to zero and by replacing 
$\a\in\Lambda$ by $i$ running from $1$ to $r$ in the results 
of the contributions from the root part. 

In the action 
$S_{VQ}^{(B)}=\int_{S^3}\L_{VQ}^{(B)}$ given in terms of the modes of 
the fluctuations, 
after completing the square by shifting the variables, 
one finds that 
\be
S_{VQ}^{(B)}
=\sum_{\a\in\Lambda_+}\sum_{l=0}^{\infty}\sum_{m=-l}^{l}\sum_{\mt=-l}^{l}
S^{(B)}_{VQ;\a,l,m,\mt}+S_{VQ;H}^{(B)},
\ne
with $\Lambda_+$ the set of the positive roots, 
where $S_{VQ;H}^{(B)}$ consists of the modes in the Cartan subalgebra of $G$.  
The modes in each $S^{(B)}_{VQ;\a,l,m,\mt}$ decouple from the modes in 
the rest of $S_{VQ}^{(B)}$. The action $S^{(B)}_{VQ;\a,l,m,\mt}$ 
of the modes for $l\ge1$, $-(l-1)\le{m}\le{l-1}$, $-l\le\mt\le{l}$, 
is given by
\be
&&S^{(B)}_{VQ;\a,l,m,\mt}=\hf\bigg[
K^\a_{l,m}g^{\bzz}\l|\t{v}^\a_{z,l,m,\mt}\r|^2
+K^\a_{l,-m}g^{\bzz}\l|\t{v}^\a_{\bz,l,m,\mt}\r|^2
+\l(\t\s^\a_{l,m,\mt}\r)^*\Delta^\a_{l,m}\t\s^\a_{l,m,\mt}
\nn\\
&&\hskip2.7cm+
\begin{array}{c}
\l(\begin{array}{ll}\l(\t{v}^\a_{l+1,m;l,\mt}\r)^*,
&\l({\t{v}}^\a_{l-1,m;l,\mt}\r)^*\end{array}\r)
\\
\\
\end{array}
\l(\begin{array}{cc}A^\a_{l,m}&B^\a_{l,m}\\
C^\a_{l,m}&D^\a_{l,m}\end{array}\r)
\l(\begin{array}{l}\t{v}^\a_{l+1,m;l,\mt}\\ 
{\t{v}}^\a_{l-1,m;l,\mt}\end{array}\r)
\bigg],
\ne
where the above operators are defined by
\be
&&K_{l,m}^\a=\l[4l(l+1)-4m+g^2\as^2\r], 
\nn\\
&&\Delta^\a_{l,m}=-{4(l-m)(l+m+1)\over{K}^\a_{l,m}}g^{\bzz}D_\bz{D}_z
-{4(l+m)(l-m+1)\over{K}^\a_{l,-m}}g^{\bzz}D_z{D}_\bz+4l(l+1),
\nn\\
&&A^\a_{l,m}=\,U^\a_{l,m}{1\over\Delta^\a_{l,m}}V^\a_{l,m}+u_{l,m}v_{l,m},
\qquad
B^\a_{l,m}=\,U^\a_{l,m}{1\over\Delta^\a_{l,m}}\t{V}^\a_{l,m}
+u_{l,m}\t{v}_{l,m},
\nn\\
&&C^\a_{l,m}=\,\t{U}^\a_{l,m}{1\over\Delta^\a_{l,m}}V^\a_{l,m}
+\t{u}_{l,m}v_{l,m},
\qquad
D^\a_{l,m}=\,\t{U}^\a_{l,m}{1\over\Delta^\a_{l,m}}\t{V}^\a_{l,m}
+\t{u}_{l,m}\t{v}_{l,m},
\ne
with 
\be
&&U^\a_{l,m}
=2(l+m)\l[2(l+1)+ig\as\r]\,\sqrt{{(l+1)(l+m+1)\over(2l+1)(l-m+1)}}
\l[{2\over{K}^\a_{l,-m}}g^{\bzz}D_z{D}_\bz-{l\over{l+m}}\r],
\nn\\
&&\t{U}^\a_{l,m}=2(l+m+1)\l[2l-ig\as\r]\,\sqrt{{l(l+m)\over(2l+1)(l-m)}}
\l[{2\over{K}^\a_{l,-m}}g^{\bzz}D_z{D}_\bz-{l+1\over{l-m+1}}\r],
\nn\\
&&V^\a_{l,m}=2(l-m)\l[2\l(l+1\r)-ig\as\r]\,
\sqrt{{(l+1)(l-m+1)\over(2l+1)(l+m+1)}}
\l[{2\over{K}^\a_{l,m}}g^{\bzz}D_\bz{D}_z-{l\over{l-m}}\r],
\nn\\
&&\t{V}^\a_{l,m}=2(l-m+1)\l[2l+ig\as\r]\,\sqrt{{l(l-m)\over(2l+1)(l+m)}}
\l[{2\over{K}^\a_{l,m}}g^{\bzz}D_\bz{D}_z-{l+1\over{l+m+1}}\r],
\ne
\be
&&u_{l,m}=\sqrt{{(l+1)(l-m+1)\over(2l+1)(l+m+1)}}\l[2(l+1)+ig\as\r],
\nn\\
&&v_{l,m}=\sqrt{{(l+1)(l+m+1)\over(2l+1)(l-m+1)}}\l[2(l+1)-ig\as\r],
\nn\\
&&\t{u}_{l,m}=-\sqrt{{l(l+m)\over(2l+1)(l-m)}}\l[2l-ig\as\r],
\qquad
\t{v}_{l,m}=-\sqrt{{l(l-m)\over(2l+1)(l+m)}}\l[2l+ig\as\r].
\ne
Note here that the covariant derivatives $D_\bz$, $D_z$ acting 
on the root part of a field, $\Phi^\a$ gives 
\be
D_\bz\Phi^\a=\d_\bz\Phi^\a+ig\sum_{i=1}^{r}\a_i{v}^i_\bz\Phi^\a,
\qquad
D_z\Phi^\a=\d_z\Phi^\a+ig\sum_{i=1}^{r}\a_i{v}^i_z\Phi^\a.
\ne
One can now see that the one-loop determinant from these modes yields 
\be
&&\prod_{\a\in\Lambda_+}\prod_{l=1}^{\infty}\prod_{m=-(l-1)}^{l-1}
\prod_{\mt=-l}^{l}\bigg[{1\over\Det{1,0}{K^\a_{l,m}}\Det{0,1}{K^\a_{l,-m}}
\Det{0,0}{\Delta^\a_{l,m}}}
\nn\\
&&\hskip4.0cm
\times{\Det{0,0}{\Delta^\a_{l,m}}\over
\Det{0,0}{\t{u}_{l,m}U^\a_{l,m}-{u}_{l,m}\t{U}^\a_{l,m}}
\Det{0,0}{\t{v}_{l,m}V^\a_{l,m}-{v}_{l,m}\t{V}^\a_{l,m}}}
\bigg]
\nn\\
&&=\prod_{\a\in\Lambda_+}\prod_{l=1}^{\infty}\prod_{m=-(l-1)}^{l-1}
\prod_{\mt=-l}^{l}\bigg[
{1\over\Det{0,0}{4l(l+1)}}
{\Det{0,0}{K^\a_{l,m}}\Det{0,0}{K^\a_{l,-m}}\over
\Det{1,0}{K^\a_{l,m}}\Det{0,1}{K^\a_{l,-m}}}
\nn\\
&&\hskip4.2cm
\times{
\Det{0,0}{4l^2+g^2\as^2}\Det{0,0}{4(l+1)^2+g^2\as^2}
\over\Det{0,0}{2g^{\bzz}D_\bz{}D_z-K^\a_{l,m}}
\Det{0,0}{2g^{\bzz}D_z{}D_\bz-K^\a_{l,-m}}}
\bigg]
\nn\\
&&=\prod_{\a\in\Lambda}\prod_{l=1}^{\infty}\prod_{m=-(l-1)}^{l-1}
\prod_{\mt=-l}^{l}\bigg[
{\Det{0,0}{K^\a_{l,m}}\over
\Det{1,0}{K^\a_{l,m}}}{1\over\Det{0,0}{2g^{\bzz}D_\bz{}D_z-K^\a_{l,m}}}
\nn\\
&&\hskip4.2cm
\times{
\Det{0,0}{2(l+1)+ig\as}\Det{0,0}{-2l+ig\as}
\over\sqrt{\Det{0,0}{4l(l+1)}}}
\bigg].
\ne

Further, after some similar algebra,  the action $S^{(B)}_{VQ;\a,l,m,\mt}$ of 
the modes for $l\ge{1/2}$, $m=-{l}$, $-l\le\mt\le{l}$ 
can be read as 
\be
\hf\bigg[
K^\a_{l,-l}g^{\bzz}\l|\t{v}^\a_{z,l,-l,\mt}\r|^2
+K^\a_{l,l}g^{\bzz}\l|\t{v}^\a_{\bz,l,-l,\mt}\r|^2
+{4l\over{K}^\a_{l,-l}}\l(\t\s^\a_{l,-l,\mt}\r)^*
\l[-2g^{\bzz}D_\bz{D}_z+(l+1)K^\a_{l,-l}\r]\t\s^\a_{l,-l,\mt}
\nn\\
+(l+1)\l(\t{v}^\a_{l+1,-l;l,\mt}\r)^*
{K^\a_{l,-l-1}\over-2g^{\bzz}D_\bz{D}_z+(l+1)K^\a_{l,-l}}
\l[-2g^{\bzz}D_{\bz}D_z+K^\a_{l,-l}\r]
\t{v}^\a_{l+1,-l;l,\mt}
\bigg],
\ne
and the one for $l\ge{1/2}$, $m=+{l}$, $-l\le\mt\le{l}$ as 
\be
\hf\bigg[
K^\a_{l,l}g^{\bzz}\l|\t{v}^\a_{z,l,l,\mt}\r|^2
+K^\a_{l,-l}g^{\bzz}\l|\t{v}^\a_{\bz,l,l,\mt}\r|^2
+{4l\over{K}^\a_{l,-l}}\l(\t\s^\a_{l,l,\mt}\r)^*
\l[-2g^{\bzz}D_z{D}_\bz+(l+1)K^\a_{l,-l}\r]\t\s^\a_{l,l,\mt}
\nn\\
+(l+1)\l(\t{v}^\a_{l+1,l;l,\mt}\r)^*
{K^\a_{l,-l-1}\over-2g^{\bzz}D_z{D}_\bz+(l+1)K^\a_{l,-l}}
\l[-2g^{\bzz}D_{z}D_\bz+K^\a_{l,-l}\r]
\t{v}^\a_{l+1,l;l,\mt}
\bigg]. 
\ne
These sectors with $l\ge{1/2}$, $m=\pm{l}$, $-l\le\mt\le{l}$ gives 
the one-loop determinant
\be
&&\prod_{\a\in\Lambda_+}\prod_{l=1/2}^{\infty}\prod_{\mt=-l}^{l}
{1\over\Det{0,0}{l^2(l+1)^2}}
\l({\Det{0,0}{K^\a_{l,-l}}\over\Det{1,0}{K^\a_{l,l}}
\Det{0,1}{K^\a_{l,-l}}\Det{0,0}{K^\a_{l,-l-1}}}\r)^2
\nn\\
&&\hskip3cm
\times{1\over\Det{0,0}{-2g^{\bzz}D_{z}D_\bz+K^\a_{l,-l}}
\Det{0,0}{-2g^{\bzz}D_{\bz}D_z+K^\a_{l,-l}}}
\nn\\
&&=\prod_{\a\in\Lambda}\prod_{l=1/2}^{\infty}\prod_{\mt=-l}^{l}
{1\over\Det{0,0}{l(l+1)}}
{\Det{0,0}{K^\a_{l,-l}}\over\Det{1,0}{K^\a_{l,l}}
\Det{0,1}{K^\a_{l,-l}}\Det{0,0}{K^\a_{l,-l-1}}}
\nn\\
&&\hskip3cm
\times{1\over\Det{0,0}{-2g^{\bzz}D_{\bz}D_z+K^\a_{l,-l}}}.
\ne

Since the actions $S^{(B)}_{VQ;\a,l,\pm(l+1),\mt}$ 
of the modes with $l\ge0$ take simple forms, 
we will give the sum
\be
&&S^{(B)}_{VQ;\a,l,l+1,\mt}+S^{(B)}_{VQ;\a,l,-(l+1),\mt}
\nn\\
&&=
\hf\l(\t{v}^\a_{l+1,l+1;l,\mt}\r)^*
\l[-2g^{\bzz}D_z{D}_\bz+K^\a_{l,-l-1}\r]
\t{v}^\a_{l+1,l+1;l,\mt}
\nn\\
&&\quad+
\hf\l(\t{v}^\a_{l+1,-(l+1);l,\mt}\r)^*
\l[-2g^{\bzz}D_\bz{D}_z+K^\a_{l,-l-1}\r]
\t{v}^\a_{l+1,-(l+1);l,\mt},
\ne
to yield the one-loop determinant
\be
\prod_{\a\in\Lambda_+}\prod_{l=0}^{\infty}\prod_{\mt=-l}^{l}
{1\over\Det{0,0}{-2g^{\bzz}D_z{D}_\bz+K^\a_{l,-l-1}}
\Det{0,0}{-2g^{\bzz}D_\bz{D}_z+K^\a_{l,-l-1}}}. 
\ne

Finally, one can find the action $S^{(B)}_{VQ;\a,0,0,0}$  
\be
\hf\bigg[
K^\a_{0,-1}\l|\t{v}^\a_{1,0;0,0}\r|^2
+g^{\bzz}\l|
D_z\t{v}^\a_{1,0;0,0}
+g\as\t{v}^\a_{z,0,0,0}\r|^2
+g^{\bzz}\l|D_\bz\t{v}^\a_{1,0;0,0}
-g\as\t{v}^\a_{\bz,0,0,0}\r|^2
\bigg], 
\ne
of the modes with $l=0$, and it gives the one-loop determinant 
\be
\prod_{\a\in\Lambda_+}{1\over\Det{0,0}{K^\a_{0,-1}}\Det{1,0}{g\as}
\Det{0,1}{g\as}}. 
\ne

Let us turn to the one-loop determinant from the fermionic fluctuations. 
To this end, we will identify the spin operator ${\cal S}^a$ 
($a=1,2,3$) with the gamma matrix $(1/2)\g^a$ ($a=1,2,3$), respectively, 
and one can easily verify that they obeys the $SU(2)$ algebra 
\be
\l[{\cal S}^a,{\cal S}^b\r]=i\e^{abc}{\cal S}^c.
\ne
One can easily see that the left-invariant vector fields 
$L_a=-(i/2)e_a{}^m\cd_m$ 
($a=1,2,3$) also satisfy the $SU(2)$ algebra 
\be
\l[L_a,\,L_b\r]=i\e_{abc}L_c.
\ne
Therefore, on the spinors $\t\la$, $\t\psi$, one finds that
\be
\g^m\cd_m\t\la=\g^a\l(2iL_a+{1\over4}\l(\om_a\r)^{bc}\g^{bc}\r)\t\la
=2i\l[\l(L_a+{\cal S}_a\r)^2-L_aL_a\r]\t\la,
\nn\\
\g^m\cd_m\t\psi=\g^a\l(2iL_a+{1\over4}\l(\om_a\r)^{bc}\g^{bc}\r)\t\psi
=2i\l[\l(L_a+{\cal S}_a\r)^2-L_aL_a\r]\t\psi. 
\ne

In order to obtain the spherical harmonics expansion of the spinors 
$\t\la$, $\t\psi$, it is useful to introduce the eigenspinors 
$\eta_{J,M;l,\mt}$ of the operator $\g^m\cd_m$ by
\be
\eta_{J,M;l,\mt}=\sum_{m=-l}^{l}\sum_{s=\pm(1/2)}
\langle l, m; \hf, s | J, M \rangle\rangle \vp_{l,m,\mt}\zeta'_s,
\ne
with $\langle l, m; \hf, s | J, M \rangle\rangle$ 
the Clebsch-Gordan coefficients of the spin $l$ representation and 
the spin $1/2$ representation into the spin $J=l\pm1/2$ representation, 
where the spinors $\zeta'_\pm$ satisfy that 
${\cal S}^3\zeta'_\pm=\pm(1/2)\zeta'_\pm$. 

They have their eigenvalues 
\be
&&\g^m\cd_m\,\eta_{l+\hf,m+\hf; l,\mt}
=i(2l+{3\over2})\,\eta_{l+\hf,m+\hf; l,\mt},
\nn\\
&&\g^m\cd_m\,\eta_{l-\hf,m+\hf; l,\mt}
=-i(2l+{1\over2})\,\eta_{l-\hf,m+\hf; l,\mt},
\ne
and form the orthonormalized basis 
\be
\int_{S^3} \l(\eta_{J',M';l',\mt'}\r)^\dag\,\eta_{J,M;l,\mt}\,*1
=\dl_{J,J'}\dl_{M,M'}\dl_{l,l'}\dl_{m,m'}.
\ne

Substituting the spherical harmonics expansion of the spinors 
$\t\la$, $\t\psi$,
\be
&&\la=
\sum_{l=0}\sum_{m=-l-1}^{l}\sum_{\mt=-l}^{l}
\la_{l+\hf,m+\hf;l,\mt}\eta_{l+\hf,m+\hf;l,\mt}
+
\sum_{l=1/2}\sum_{m=-l}^{l-1}\sum_{\mt=-l}^{l}
\la_{l-\hf,m+\hf;l,\mt}\eta_{l-\hf,m+\hf;l,\mt},
\nn\\
&&\psi=
\sum_{l=0}\sum_{m=-l-1}^{l}\sum_{\mt=-l}^{l}
\psi_{l+\hf,m+\hf;l,\mt}\eta_{l+\hf,m+\hf;l,\mt}
+
\sum_{l=1/2}\sum_{m=-l}^{l-1}\sum_{\mt=-l}^{l}
\psi_{l-\hf,m+\hf;l,\mt}\eta_{l-\hf,m+\hf;l,\mt},
\ne
into the Lagrangian $\L_{VQ}^{(F)}$, one obtains 
the action $S_{VQ}^{(F)}=\int_{S^3}\L_{VQ}^{(F)}*1$.
There one finds the terms including  
\be
\int_{S^3} \l(\eta_{J',M';l',\mt'}\r)^\dag
\l(1-k_m\g^m\r)\eta_{J,M;l,\mt}\,*1
=\langle\langle{J',M'}|1-2{\cal S}^3|{J,M}\rangle\rangle\,\dl_{l,l'}\dl_{m,m'},
\ne
with our choice $k^a=\dl^a_3$. For the coefficients 
$\langle\langle{J',M'}|1-2{\cal S}^3|{J,M}\rangle\rangle$, 
see appendix \ref{ClebschGordan}. 

Similarly to the bosonic part, the modes $\la_{l\pm\hf,m+\hf;l,\mt}$, 
$\psi_{l\pm\hf,m+\hf;l,\mt}$ in each sector $(\a,l,m,\mt)$ decouple 
from the modes in the other sectors, and therefore 
the action $S_{VQ}^{(F)}$ can be divided into the actions 
$S_{VQ;\a,l,m\mt}^{(F)}$ of each sector $(l,m,\mt)$ as 
\be
S_{VQ}^{(F)}
=\sum_{\a\in\Lambda}\sum_{l=0}^{\infty}\sum_{m=-(l+1)}^{l}\sum_{\mt=-l}^{l}
S_{VQ;\a,l,m,\mt}^{(F)}+S_{VQ;H}^{(F)},
\ne
where $S_{VQ;H}^{(F)}$ consists of the modes in the Cartan subalgebra of $G$. 

After completing the square and shifting the fields properly, 
the action $S^{(F)}_{VQ;\a,l,m,\mt}$ of the modes for 
$l\ge1/2$, $-l\le{m}\le{(l-1)}$, $-l\le\mt\le{l}$, is given by
\be
\begin{array}{c}
(\begin{array}{ll}
\l({\psi}^\a_{z;l+\hf,m+\hf;l,\mt}\r)^\dag, 
& \l(\psi^\a_{z;l-\hf,m+\hf;l,\mt}\r)^\dag
\end{array})
\\
\quad 
\end{array}
g^{\bzz}{\cal K}^\a_{l,m}
\l(\begin{array}{l}
\psi^\a_{z;l+\hf,m+\hf;l,\mt} \\ \psi^\a_{z;l-\hf,m+\hf;l,\mt}
\end{array}\r)
\nn\\
+
\begin{array}{c}
(\begin{array}{ll}
\l(\la^\a_{l+\hf,m+\hf;l,\mt}\r)^\dag, & \l(\la^\a_{l-\hf,m+\hf;l,\mt}\r)^\dag
\end{array})
\\
\quad 
\end{array}
{\cal M}^\a_{l,m}
\l(\begin{array}{l}
\la^\a_{l+\hf,m+\hf;l,\mt} \\ \la^\a_{l-\hf,m+\hf;l,\mt}
\end{array}\r),
\ne
where
\be
&&{\cal K}^\a_{l,m}
=\l(\begin{array}{cc}
-{2l(l+1)-m\over{2l+1}}+{i\over2}g\as & -{1\over2l+1}\sqrt{(l+m+1)(l-m)} \\
-{1\over2l+1}\sqrt{(l+m+1)(l-m)} & {2l(l+1)-m\over{2l+1}}+{i\over2}g\as 
\end{array}\r),
\nn\\
&&{\cal M}^\a_{l,m}
=
\l(\begin{array}{cc}
\l(2(l+1)+ig\as\r)
& 
\\
&
\l(-2l+ig\as\r)
\end{array}\r)
\nn\\
&&\hskip2cm
\times\l(\begin{array}{cc}
-{l-m\over2l+1}{2\over{K}^\a_{l,m}}g^{\bzz}D_\bz{D}_z+1
& 
-{\sqrt{(l+m+1)(l-m)}\over2l+1}{2\over{K}^\a_{l,m}}
g^{\bzz}D_\bz{D}_z
\\
-{\sqrt{(l+m+1)(l-m)}\over2l+1}{2\over{K}^\a_{l,m}}g^{\bzz}D_\bz{D}_z
&
-{l+m+1\over2l+1}{2\over{K}^\a_{l,m}}g^{\bzz}D_\bz{D}_z+1
\end{array}\r),
\ne
and one can see that to the one-loop determinant, they yield the 
contributions
\be
&&\prod_{\a\in\La}\prod_{l=1/2}^{\infty}\prod_{m=-l}^{l-1}\prod_{\mt=-l}^{l}
{\Det{1,0}{K^\a_{l,m}}\over\Det{0,0}{K^\a_{l,m}}}
\Det{0,0}{2g^{\bzz}D_\bz{D}_z-K^\a_{l,m}}
\nn\\
&&\hskip3.3cm\times
\Det{0,0}{\l(2(l+1)+ig\as\r)\l(-2l+ig\as\r)},
\ne
up to an overall normalization constant.

For the remaining fermionic modes with $l\ge0$, $m=-(l+1),l$; $-l\le\mt\le{l}$, 
after some similar algebra, one obtains 
\be
&&\l(\la^\a_{l+\hf,l+\hf;l,\mt}\r)^\dag
\l[2(l+1)+ig\as\r]
\la^\a_{l+\hf,l+\hf;l,\mt}
\nn\\
&&+\l(\la^\a_{l+\hf,-l-\hf;l,\mt}\r)^\dag
{1\over-2(l+1)+ig\as}\l[2g^{\bzz}D_\bz{D}_z-K^\a_{l,-l-1}\r]
\la^\a_{l+\hf,-l-\hf;l,\mt}
\nn\\
&&+\hf{g}^{\bzz}
\l(\psi^\a_{z;l+\hf,l+\hf;l,\mt}\r)^\dag
\l[-2l+ig\as\r]
\psi^\a_{z;l+\hf,l+\hf;l,\mt}
\nn\\
&&+\hf{g}^{\bzz}
\l(\psi^\a_{z;l+\hf,-l-\hf;l,\mt}\r)^\dag
\l[-2(l+1)+ig\as\r]
\psi^\a_{z;l+\hf,-l-\hf;l,\mt}, 
\ne
and finds the one-loop determinant 
\be
&&\prod_{\a\in\Lambda}\prod_{l=0}^{\infty}\prod_{\mt=-l}^{l}
\Det{1,0}{2(l+1)+ig\as}\Det{1,0}{-2l+ig\as}
\nn\\
&&\hskip7cm
\times
\Det{0,0}{-2g^{\bzz}D_\bz{D}_z+K^\a_{l,-l-1}},
\ne
up to an overall constant.

Wrapping up the contributions from the bosonic fluctuations and 
the fermionic fluctuations to the one-loop determinant, 
one obtains 
\be
\prod_{\a\in\Lambda_+}{1\over\Det{1,0}{K^\a_{0,0}}}
\prod_{l=\hf}^{\infty}
\l({\Det{0,0}{K^\a_{l,l}}\over\Det{1,0}{K^\a_{l,l}}}\r)^2,
\ne
up to an overall normalization constant.
Taking account of this result and the one-loop determinant (\ref{DetFP}) 
from the ghost, and using the same reason of the Hodge decomposition 
as in \cite{Lecture}, one finds that the total one-loop determinant 
is given by 
\be
\l(\prod_{\a\in\Lambda_+}\sin\l(i{\pi}g\as\r)\r)^{\chi(\Sigma)}. 
\label{mainresult}
\ee
This is one of the main results in this paper. 

There are two subtle points related to the zero modes in the Cartan 
subalgebra, on which so far we have not discussed in detail.  
In the Cartan part, there is the fermion zero modes 
$(\psi^i_{z;\hf,\hf,0,0})^\dag$, $\psi^i_{z;\hf,\hf,0,0}$, 
but the term $g^{\bzz}\bar\psi_\bz\psi_z$ in the Lagrangian $\L'_V$ 
absorbs them. Therefore, they cause no problems. 

We are left to perform the path integral over the background gauge fields 
$v_z^i(z,\bz)$ ($i=1,\cdots,r$), along with the finite-dimensional 
integral over the background $\s^i$ ($i=1,\cdots,r$). 
While the background of the gauge fields $v_z^i(z,\bz)$ obeying 
$
\int_{\Sigma} v^i_{\bzz}d\bz\wedge{dz}\not=0
$
in the classical action $\L_{\rm cl}$ can contribute to the path integral,
Upon the integration over the gauge fields $v_z^i(z,\bz)$, 
the fluctuations of the gauge fields around the background don't appear 
in the rest of the path integral. One therefore needs to divide 
the integration over the fluctuations, 
along with the other possible constant factors.

\subsection{The Contribution from the Hypermultiplet}
\label{LocalizationH}

Let us proceed to the hypermultiplet. 
Along with the BRST transformation (\ref{BRSTV}) of the vector multiplet, 
the hypermultiplet transform under the BRST transformation as
\be
&&\dl_Q\t{H}=0, 
\qquad
\dl_Q{H}=0, 
\nn\\
&&\dl_Q\l(\t{H}\r)^*=-i\l(\t\chi\r)^\dag\e,
\qquad
\dl_Q\l({H}\r)^*=-i\l(\chi\r)^\dag\e,
\nn\\
&&\dl_Q\t{\chi}=\l[D_m\t{H}\g^m-g\l[\s,\,\t{H}\r]+i\t{H}\r]\e,
\qquad
\dl_Q\l(\t{\chi}\r)^\dag=\e^TC_3\l[2iD_zH+\l(F_{H2}\r)^*\r],
\nn\\
&&\dl_Q{\chi}=\l[D_m{H}\g^m-g\l[\s,\,{H}\r]+i{H}\r]\e,
\qquad
\dl_Q\l({\chi}\r)^\dag=-\e^TC_3\l[2iD_z\t{H}+F_{H1}\r],
\nn\\
&&\dl_QF_{H1}=0,
\qquad
\dl_Q\l(F_{H2}\r)^*=0,
\label{BRSTH}\\
&&\dl_Q\l(F_{H1}\r)^*
=i\bigg[-D_m\chi^TC_3\g^m-2i\l(D_z\t{\chi}\r)^\dag-{i\over2}\chi^TC_3
\nn\\
&&\hskip5cm
+g\l[\s,\,\chi^T\r]C_3-2ig\l[\t{H}^*,\,\psi^\dag\r]-2i\l[H,\,\la^T\r]C_3
\bigg]\e,
\nn\\
&&\dl_QF_{H2}
=i\e^TC_3\bigg[\g^mD_m\t\chi-2iC_3^{-1}\l(D_z{\chi}\r)^*-{i\over2}\t\chi
\nn\\
&&\hskip5cm
+g\l[\s,\,\t\chi\r]-2ig\l[\t{H},\,\la\r]-2iC_3^{-1}\l[H^*,\,\psi^*\r]
\bigg].
\ne
Note that the analytic continuation for the scalar field $\s$ has 
already been done here.  

In order to carry out the localization procedure, 
we will add the Lagrangian 
\be
\L_{HQ}=\dl_Q\l[
\l(\dl_Q\t\chi\r)^\dag\t\chi+\l(\t\chi\r)^\dag\l(\dl_Q\l(\t\chi\r)^\dag\r)^\dag
+\l(\dl_Q\chi\r)^\dag\chi+\l(\chi\r)^\dag\l(\dl_Q\l(\chi\r)^\dag\r)^\dag
\r],
\label{LHQ}
\ee
to the Lagrangian $\L_{VQ}$ in (\ref{LVQ}). 
The total Lagrangian $\L$ will thus be shifted as 
$\L\to\L-t\l(\L_{VQ}+\L_{HQ}\r)$.

A fixed point is given by a solution to $\dl_Q\chi=0$, $\dl_Q\t\chi=0$ 
meaning that 
\be
D_m\t{H}\,\g^m+i\t{H}-g\l[\s,\,\t{H}\r]=0,
\qquad
D_m{H}\,\g^m+i{H}-g\l[\s,\,{H}\r]=0,
\ne
and to $\dl_Q{\t\chi}^\dag=0$, $\dl_Q{\t\chi}^\dag=0$. 
Since the solution to the former equations is given by 
$\t{H}=0$, $H=0$, substituting it into the latter equations, 
one obtains the solution $F_{H1}=0$, $F_{H2}=0$. 
One thus finds no non-trivial backgrounds. 

Then, up to quadratic order of the fluctuations, 
the bosonic part $\L_{HQ}^{(B)}$ of the Lagrangian $\L_{HQ}$ is given by 
\be
&&{\rm tr}\bigg[
\l(D_m\t{H}\r)^\dag{D}^m\t{H}+\l(D_m{H}\r)^\dag{D}^m{H}
\nn\\
&&\quad
+\l(\t{H}+ig\l[\s,\,\t{H}\r]\r)^\dag\l(\t{H}+ig\l[\s,\,\t{H}\r]\r)
+\l({H}+ig\l[\s,\,{H}\r]\r)^\dag\l({H}+ig\l[\s,\,{H}\r]\r)
\nn\\
&&\quad
+\l(F_{H1}+2iD_z\t{H}\r)^\dag\l(F_{H1}+2iD_z\t{H}\r)
+\l(F_{H2}-2i\l(D_z\t{H}\r)^*\r)^\dag\l(F_{H2}-2i\l(D_z\t{H}\r)^*\r)
\bigg],
\ne
where $\s$ is the fixed point (\ref{Cartansigma}), 
and the fermionic part $\L_{HQ}^{(F)}$ by
\be
{\rm tr}\bigg[
\t\chi^\dag\,k_n\g^n\l(i\g^mD_m\t\chi+\hf\t\chi-ig\l[\s,\,\t\chi\r]\r)
+
\chi^\dag\,k_n\g^n\l(i\g^mD_m\chi+\hf\chi-ig\l[\s,\,\chi\r]\r)
\bigg]. 
\ne

We will carry out similar calculations to what we have done 
for the vector multiplet by substituting the spherical harmonic 
expansions of the fluctuations 
\be
&&\t{H}=\sum_{l=0}^{\infty}\sum_{m=-l}^{l}\sum_{\mt=-l}^{l}
\t{H}_{l,m,\mt}\,\vp_{l,m,\mt},
\qquad
H=\sum_{l=0}^{\infty}\sum_{m=-l}^{l}\sum_{\mt=-l}^{l}
H_{l,m,\mt}\,\vp_{l,m,\mt},
\nn\\
&&\t{\chi}=\sum_{l=0}^{\infty}\sum_{m=-l-1}^{l}\sum_{\mt=-l}^{l}
\t\chi_{l+\hf,m+\hf;l,\mt}\,\eta_{l+\hf,m+\hf;l,\mt}
+\sum_{l=1/2}^{\infty}\sum_{m=-l}^{l-1}\sum_{\mt=-l}^{l}
\t\chi_{l-\hf,m+\hf;l,\mt}\,\eta_{l-\hf,m+\hf;l,\mt},
\nn\\
&&{\chi}=\sum_{l=0}^{\infty}\sum_{m=-l-1}^{l}\sum_{\mt=-l}^{l}
\chi_{l+\hf,m+\hf;l,\mt}\,\eta_{l+\hf,m+\hf;l,\mt}
+\sum_{l=1/2}^{\infty}\sum_{m=-l}^{l-1}\sum_{\mt=-l}^{l}
\chi_{l-\hf,m+\hf;l,\mt}\,\eta_{l-\hf,m+\hf;l,\mt},
\ne
into $\L_{HQ}$. Recalling that $k_n\g^n=2{\cal S}_3$ and using 
the Clebsch-Gordan coefficients in appendix \ref{ClebschGordan}, one obtains 
the root part of $\L_{HQ}^{(B)}$
\be
\sum_{\a\in\Lambda}~\sum_{l=0}^{\infty}\sum_{m=-l}^{l}\sum_{\mt=-l}^{l}
\l[4l(l+1)+1+g^2\as^2\r]\l(\l|\t{H}_{l,m,\mt}^\a\r|^2+
\l|{H}_{l,m,\mt}^\a\r|^2\r),
\label{LHQB}
\ee
and the root part of $\L_{HQ}^{(F)}$ is given by the sum of 
\be
&&\sum_{\a\in\Lambda}\bigg\{
\sum_{l=0}^{\infty}\sum_{\mt=-l}^{l}\bigg[
-\t\chi_{l+\hf,l+\hf;l,\mt}^\dag
\l[(2l+1)+ig\as\r]
\t\chi_{l+\hf,l+\hf;l,\mt}
\nn\\
&&\hskip 3cm
+\t\chi_{l+\hf,-l-\hf;l,\mt}^\dag
\l[(2l+1)+ig\as\r]
\t\chi_{l+\hf,-l-\hf;l,\mt}\bigg]
\label{LHQF}\\
&&-\sum_{l=1/2}^{\infty}\sum_{m=-l}^{l-1}\sum_{\mt=-l}^{l}
\bigg[
\begin{array}{l}
\l(\t\chi_{l+\hf,m+\hf;l,\mt}^\dag,\t\chi_{l-\hf,m+\hf;l,\mt}^\dag\r)
\\
\qquad
\end{array}
\l(
\begin{array}{cc}
{\cal A}^\a_{l,m} & {\cal B}^\a_{l,m} \\
{\cal C}^\a_{l,m} & {\cal D}^\a_{l,m} \\
\end{array}
\r)
\l(\begin{array}{l}
\t\chi_{l+\hf,m+\hf;l,\mt}
\\
\t\chi_{l-\hf,m+\hf;l,\mt}
\end{array}\r)
\bigg]
\bigg\},
\ne
and the same terms with $\t\chi$'s replaced by $\chi$'s,
where 
\be
&&{\cal A}^\a_{l,m}={2m+1\over2l+1}\l((2l+1)+ig\as\r),  
\nn\\
&&{\cal B}^\a_{l,m}=-2{\sqrt{(l+m+1)(l-m)}\over2l+1}\l((2l+1)+ig\as\r),
\nn\\
&&{\cal C}^\a_{l,m}=2{\sqrt{(l+m+1)(l-m)}\over2l+1}\l((2l+1)-ig\as\r),
\nn\\
&&{\cal D}^\a_{l,m}={2m+1\over2l+1}\l((2l+1)-ig\as\r).   
\ne
Note that the integration over the auxiliary fields $F_{H1}$, $F_{H2}$ 
has already been done, and their contributions to the partition function 
is just an overall constant. 

Although the Cartan part of $\L_{HQ}$ can be easily obtained by setting 
$\a$ to zero and by replacing the sum over the $\Lambda$ by 
over $i$ running from $1$ to $r$, we aren't interested in 
the overall normalization constant of the partition function, 
to which the Cartan part can only contribute. 
Therefore, we will focus on the root part, as for the vector multiplet. 

From (\ref{LHQB}) and (\ref{LHQF}), one can easily see that 
the one-loop determinant from the bosonic fluctuations yields 
\be
\prod_{\a\in\Lambda_+}\prod_{l=0}^{\infty}\prod_{m=-l}^{l}\prod_{\mt=-l}^{l} 
\l({1\over\Det{\hf,0}{(2l+1)^2+g^2\as^2}}\r)^2,
\ne
and the one from the fermionic fluctuations, 
\be
\prod_{\a\in\Lambda_+}\prod_{l=0}^{\infty}\prod_{m=-l}^{l}\prod_{\mt=-l}^{l} 
\l(\Det{\hf,0}{(2l+1)^2+g^2\as^2}\r)^2,
\ne
up to an overall constant. 

Wrapping up them, one finds that the hypermultiplet contributes 
just a constant to the total partition function. 

\section{Discussions}
\label{Discussions}

From (\ref{mainresult}) and (\ref{YM}), the partition function of 
both of the $\N=1$ and $\N=2$ theories reduce to the finite-dimensional 
integral 
\be
Z_{5DSYM}={\N}_{5DSYM}\,\sum_{m}\prod_{i=1}^{r}\int\,d\s^i 
\l[\prod_{\a\in\Lambda_+}2\sin\l(i{\pi}g\as\r)\r]^{\chi(\Sigma)}
\exp\l[\int\L_{YM}d({\rm vol})\r], 
\label{Z5DSYM}
\ee
with $m$ the first Chern number of the two-dimensional gauge field $v_z$ on 
$\Sigma$, where the normalization constant ${\N}_{5DSYM}$ may be different 
in the $\N=2$ theory from in the $\N=1$ theory. 

Let us find the parameter $q$ from (\ref{Z5DSYM}), and 
furthermore, for the comparison\footnote{
We would like to thank Yuji Tachikawa for suggesting us to check the 
consistency of our results with the conformal index in \cite{Rastelli}.} 
with the prediction from the conjecture 
\cite{Douglas,Lambert} for the six-dimensional $\N=(2,0)$ theory, 
we will replace the radius of the unit $S^3$ by $l$. 
For brevity, we will take the $SU(2)$ gauge group. 

Then, the result (\ref{mainresult}) 
\be
\l[\prod_{\a\in\Lambda_+}2\sin\l(i{\pi}g\as\r)\r]^{\chi(\Sigma)} 
\ne
reduces into
\be
\l[2\sin\l(i\sqrt{2}{\pi}gl\s\r)\r]^{\chi(\Sigma)}. 
\label{5Dmeasure}
\ee

From the Lagrangian (\ref{YM})
\be
\L_{YM}=2\pi^2l^3\tr{-\l({\s\over{l}}\r)^2+2{\s\over{l}}\,{g}^{\bzz}v_{\bzz}},
\ne
the classical action is given by
\be
\int_\Sigma \L_{YM}\, d({\rm vol})
=-2\pi^2l\,\int_\Sigma d(vol)\l(\s\r)^2
-4i\pi^2l^2\s\int_\Sigma v_{\bzz}d\bz\wedge{d}z.
\ne
As explained in detail in \cite{Lecture}, we need the summation over 
the first Chern numbers of the two-dimensional gauge field $v_z$, $v_\bz$ 
on $\Sigma$. Here, let us explain the fact that 
the normalization of the first Chern number is given by 
\be
\int_\Sigma v_{\bzz}d\bz\wedge{d}z={\sqrt{2}\over{g}}\, 2\pi{m}, 
\label{monopole}
\ee
with $m\in\IZ$. 
Let us recall that the Cartan subalgebra of the $SU(2)$ gauge group 
is generated by $H=\s_3/\sqrt{2}$, 
with the normalization $\tr{HH}=1$. The Lie algebra of the $SU(2)$ gauge group 
in fact is generated by $H$ and $E_{\pm}$, which obey that
\be
\l[H,\, E_{\pm}\r]=\pm\sqrt{2}E_{\pm}, 
\qquad
\l[E_{+},\, E_{-}\r]=\sqrt{2}H, 
\ne
in our convention. Therefore, a field $\psi$ in the fundamental 
representation of the gauge group can be decomposed into the eigenstates 
of $H$ as
\be
H\psi_{\pm}=\pm{1\over\sqrt{2}}\,\psi_{\pm}, 
\ne
and the covariant derivative gives 
\be
D\psi_+
=d\psi_++{i\over\sqrt{2}}gv\psi_{+}.
\ne
Under a gauge transformation, the two-dimensional gauge field $v_z$ transforms 
in a differential form notation as
\be
v \quad\to\quad v - {\sqrt{2}\over{g}}d\Omega, 
\ne
and then the field $\psi_+$ transforms as 
\be
\psi_+ \quad\to\quad e^{i\Omega}\psi_+.
\ne
For brevity, let us take $\Sigma=S^2$ and consider two patches 
$U_N=\{(\theta,\phi)|0\le\theta\le{\pi/2}\}$ and 
$U_S=\{(\theta,\phi)|{\pi/2}\le\theta\le{\pi}\}$, 
covering the $S^2$ with the polar coordinates ($\theta$, $\phi$). 
On $U_N\cap{U}_S$, suppose that the section $\psi_N$ of $\psi_+$ on $U_N$ is 
related to the section $\psi_S$ on $U_S$ as $\psi_S=e^{i\Omega}\psi_N$. 
Then, the requirement that $\psi_S$ be single-valued is satisfied if 
\be
\Omega=m\,\phi, \qquad (m \in \IZ)
\ne
on the $U_N\cap{U}_S$. Then, the flux is determined as
\be
\int_{\Sigma}dv=\int_{U_N\cap{U}_S}\l(v_N-v_S\r)
=\int_{U_N\cap{U}_S}{\sqrt{2}\over{g}}d\Omega={\sqrt{2}\over{g}}(2\pi{m}). 
\ne

Substituting the gauge field configurations (\ref{monopole}) 
into the partition function (\ref{Z5DSYM}) 
and summing up over the Chern number $m$, 
one can see that the dominant contribution from 
the integration over $\s$ is given by the points 
\be
\s={g\over\sqrt{2}}{n\over4\pi^2{l}^2},
\ne
with $n\in \IZ$.
Since the measure (\ref{5Dmeasure}) at the dominant points of $\s$ yields
\be
\l(2\sin\l(i{g^2\over4\pi{l}}n\r)\r)^{\chi(\Sigma)}
=\l[e^{-{g^2\over4\pi{l}}\cdot{n}}-e^{+{g^2\over4\pi{l}}\cdot{n}}
\r]^{\chi(\Sigma)}
=\l[n\r]_q,
\ne
we obtain the parameter 
\be
q=\exp(-{g^2\over2\pi{l}}).
\label{5Dq}
\ee

In the paper \cite{Rastelli}, the superconformal index in 
four-dimensional $\N=2$ gauge theories was calculated on $S^3\times{S}^1$. 
In the index, one can see that the parameter $q$ is found in the form 
\be
q^{\Delta}=e^{-\beta{E}},
\ne
where $\Delta$ is the conformal weight of states over which the index 
have the summation, and the energy $E$ can be obtained through 
the state-operator mapping in conformal field theories. 
The temperature $\beta$ is the radius of $S^1$, which we 
regard  as the circle on which the six-dimensional $\N=(2,0)$ theory 
is placed to yield the five-dimensional $\N=2$ theory.

Following \cite{Douglas,Lambert}, instanton solutions 
in the five-dimensional $\N=2$ theory 
correspond to the Kaluza-Klein modes in the six-dimensional $\N=(2,0)$ theory 
compactified on $S^1$. In a four-dimensional $SU(2)$ gauge theory, 
the one-instanton solution gives the classical action
\be
\int d^4x F^a_{mn}F^a_{mn}=32\pi^2,
\ne
where $m,n=1,\cdots,4$, with the normalization 
\be
F_{mn}^a=\d_mA^a_n-\d_nA^a_m+\e^{abc}A^b_mA^c_n.
\ne
Therefore, for our convention, identifying 
\be
v_m^a=-{1\over\sqrt{2}g}A^a_m, \qquad (m=1,\cdots,4)
\ne
we can see that 
\be
v^a_{mn}=\d_mv^a_n-\d_nv^a_m-\sqrt{2}g\e^{abc}v^b_mv^c_n
=-{1\over\sqrt{2}g}F^a_{mn},
\ne
where $m,n=1,\cdots,4$, 
and thus in the five-dimensional theory, one finds the classical action 
\be
\int d^5X -{1\over4}v^a_{MN}v^a_{MN}=-{4\pi^2\over{g}^2}\int dX_5
\ne
for the instanton solution of unit instanton charge. 

For the six-dimensional $(2,0)$ theory on $S^1$ of radius $R$, 
the instanton solution of unit instanton charge 
corresponds to the first KK modes, and so one obtains the relation 
\be
{1\over{R}}={4\pi^2\over{g}^2}.
\ne
One thus finds the temperature 
\be
\beta=2\pi{R}={g^2\over2\pi}.
\label{R}
\ee

For a 4-dimensional massless scalar of conformal weight $\Delta=1$, 
the conformal coupling term of it with the scalar curvature in the Lagrangian 
gives it a mass 
\be
E=m={1\over{l}},
\ne
on $\IR\times{}S^3$, where the radius of the $S^3$ is $l$. 
Therefore, the state-operator mapping in conformal field theories suggests that 
\be
E={\Delta\over{l}}.
\label{E}
\ee

Wrapping up (\ref{R}) and (\ref{E}) to obtain 
\be
e^{-\beta{E}}=e^{-{g^2\over2\pi}{\Delta\over{l}}}
=e^{-{g^2\over2\pi{l}}\Delta},
\ne
the parameter $q$ can now be read as 
\be
q=e^{-{g^2\over2\pi{l}}},
\ne
which is in perfect agreement with our result in the five-dimensional theory. 

Thus, we have seen that the partition function of five-dimensional theory 
yields the partition function of the two-dimensional $q$-deformed Yang-Mills 
theory, but not the ordinary Yang-Mills theory on a closed Riemann surface. 
It is consistent with the proposal in \cite{Rastelli}. 

Furthermore, in order for 
the parameter $q$ found in the result of the five-dimensional theory 
to be identical to the $q$ in the conformal index of \cite{Rastelli}, 
we must identify the five-dimensional gauge coupling constant $g$ as 
the temperature $\beta$ or equivalently the radius $R$ of the $S^1$ 
in the four-dimensional theory. However, the identification is also 
consistent with the prediction of the conjecture \cite{Douglas,Lambert} 
that instanton solutions in the five-dimensional $\N=2$ theory 
is identical to the Kaluza-Klein modes in the six-dimensional $\N=(2,0)$ 
theory. 

Since the hypermultiplet gives no contributions to these results, 
what we have discussed above is also the case for the five-dimensional 
$\N=1$ theory. However, the existence of the corresponding six-dimensional 
theory is not clear for us up to this point. 

Recently, the authors of \cite{qtYM} and of \cite{pqtYM} have considered 
superconformal indices of four-dimensional $\N=2$ gauge theories 
with more parameters, and found that the indices give rise 
to not just the $q$-deformed two-dimensional Yang-Mills theory, 
but the ($q,t$)-deformed and the ($p$,$q$,$t$)-deformed ones. 
It would be interesting to extend the localization analysis in this paper 
to these cases. 

\vskip 2cm
\vskip 5cm
\centerline{\bf Acknowledgement}

\medskip

The authors would like to thank Kazuo Hosomichi for helpful discussions. 
We are also grateful to Yuji Tachikawa for helpful discussions, 
valuable suggestions and a careful reading of the manuscript. 
The work of T.~K. was supported in part 
by a Grant-in-Aid \#23540286 from the MEXT of Japan.

\break
\appendix
\section{Gamma Matrices}
\label{notations}

The five-dimensional gamma matrices $\G^M$ ($M=1,\cdots,5$) 
satisfy  
\be
\l\{\G^M,\,\G^N\r\}=2\delta^{MN}, 
\ne
and they are given in terms of the three-dimensional gamma matrices 
$\g^m=\s_m$ ($m=1,2,3$) as
\be
\G^m=\g^m\otimes \s_2, 
\qquad
\G^4=\1\otimes\s_1,
\qquad
\G^5=\1\otimes\s_3,
\ne
where $\s_{1,2,3}$ are the Pauli matrices. 

The five-dimensional charge conjugation matrix $C_5$ satisfies 
\be
\l(\G^M\r)^T=C_5\, \G^M\, C_5^{-1},
\qquad
\l(C_5\r)^T=-C_5,
\ne
where $T$ denotes the transpose of the matrix, and it 
may be given in terms of the three-dimensional charge conjugate matrix 
$C_3=i\s_2$ as 
\be
C_5=C_3\otimes\1.
\ne

\section{The Clebsch-Gordan Coefficients}
\label{ClebschGordan}

\subsection{The spin $l$ representation $\otimes$ the spin $1$ 
representation into the spin $J=l,l\pm1$ representations}

\begin{itemize}

\item the spin $J=l+1$ representation
\be
&&\l| J=l+1, M=m \r.\rangle\rangle
\nn\\
&&=\sqrt{{1\over2(l+1)(2l+1)}}\bigg[
\sqrt{(l+m)(l+m+1)}\l| l, m-1 \r.\rangle\l| 1, 1 \r.\rangle
\nn\\
&&\quad
+\sqrt{2(l+m+1)(l-m+1)}\l| l, m \r.\rangle\l| 1, 0 \r.\rangle
+\sqrt{(l-m)(l-m+1)}\l| l, m+1 \r.\rangle\l| 1, -1 \r.\rangle
\bigg].
\ne

\item the spin $J=l$ representation
\be
&&\l| J=l, M=m \r.\rangle\rangle
\nn\\
&&=\sqrt{{1\over2l(l+1)}}\bigg[
-\sqrt{(l+m)(l-m+1)}\l| l, m-1 \r.\rangle\l| 1, 1 \r.\rangle
\nn\\
&&\hskip4cm
+\sqrt{2}m\l| l, m \r.\rangle\l| 1, 0 \r.\rangle
+\sqrt{(l-m)(l+m+1)}\l| l, m+1 \r.\rangle\l| 1, -1 \r.\rangle
\bigg].
\ne

\item the spin $J=l-1$ representation
\be
&&\l| J=l-1, M=m \r.\rangle\rangle
\nn\\
&&=\sqrt{{1\over2l(2l+1)}}\bigg[
\sqrt{(l-m)(l-m+1)}\l| l, m-1 \r.\rangle\l| 1, 1 \r.\rangle
\nn\\
&&\quad
-\sqrt{2(l+m)(l-m)}\l| l, m \r.\rangle\l| 1, 0 \r.\rangle
+\sqrt{(l+m)(l+m+1)}\l| l, m+1 \r.\rangle\l| 1, -1 \r.\rangle
\bigg].
\ne

\end{itemize}

Furthermore, the action of the spin operator $S_3$ on the state 
$\l| J, M \r.\rangle\rangle$ is obtained as 
\be
&&S_3\l| l+1, M \r.\rangle\rangle
\nn\\
&&\quad={M\over{l+1}}\l| l+1, M \r.\rangle\rangle
-{1\over{l+1}}\sqrt{{l\over2l+1}}\sqrt{(l+M+1)(l-M+1)}
\l| l, M \r.\rangle\rangle,
\nn\\
&&S_3\l| l, M \r.\rangle\rangle
\nn\\
&&\quad=\sqrt{{1\over{l(l+1)}}}
\bigg[
-l\sqrt{{(l-M+1)(l+M+1)\over(l+1)(2l+1)}}\l| l+1, M \r.\rangle\rangle
\nn\\
&&\qquad
+M\sqrt{{1\over{l}(2l+1)}}\l| l, M \r.\rangle\rangle
-(l+1)\sqrt{{(l-M)(l+M)\over{l}(2l+1)}}\l| l-1, M \r.\rangle\rangle
\bigg],
\nn\\
&&S_3\l| l-1, M \r.\rangle\rangle
\nn\\
&&\quad=-{M\over{l}}\l| l-1, M \r.\rangle\rangle
-{1\over{l}}\sqrt{{l+1\over2l+1}}\sqrt{(l+M)(l-M)}
\l| l, M \r.\rangle\rangle.
\ne

\subsection{The spin $l$ representation $\otimes$ the spin $1/2$ 
representation into the spin $J=l\pm1/2$ representations}

\begin{itemize}

\item the spin $J=l+1/2$ representation 
($m=-l-1,-l,\cdots,l-1,l$)
\be
&&| J=l+\hf, M=m+\hf \rangle\rangle
=
\sqrt{{l-m\over2l+1}}\l| l, m+1 \r.\rangle \,| \hf, -\hf \rangle
+\sqrt{{l+m+1\over2l+1}}\l| l, m \r.\rangle \,| \hf, \hf \rangle. 
\ne

\item the spin $J=l-1/2$ representation 
($m=-l,\cdots,l-1$)
\be
&&| J=l-\hf, M=m+\hf \rangle\rangle
=
\sqrt{{l+m+1\over2l+1}}\l| l, m+1 \r.\rangle \,| \hf, -\hf \rangle
-\sqrt{{l-m\over2l+1}}\l| l, m \r.\rangle \,| \hf, \hf \rangle.
\ne

\end{itemize}

Furthermore, the action of the spin operator $S_3$ on the state 
$\l| J, M \r.\rangle\rangle$ is obtained as 
\be
&&S_3| l+\hf, m+\hf \rangle\rangle
\nn\\
&&\quad
=\hf{2m+1\over2l+1}\,| l+\hf, m+\hf \rangle\rangle
-{\sqrt{(l+m+1)(l-m)}\over2l+1}\,| l-\hf, m+\hf \rangle\rangle,
\nn\\
&&S_3| l-\hf, m+\hf \rangle\rangle
\nn\\
&&\quad
=-{\sqrt{(l+m+1)(l-m)}\over2l+1}\,| l+\hf, m+\hf \rangle\rangle
-\hf{2m+1\over2l+1}\,| l-\hf, m+\hf \rangle\rangle.
\ne

\clearpage

\end{document}